\pdfoutput=1
\documentclass[iop]{emulateapj}
\usepackage{longtable}
\usepackage{amsmath}
\usepackage{amssymb}
\usepackage{amsthm}
\usepackage{graphicx}
\shorttitle{ Geometry of Multi-Transiting Systems }
\shortauthors{Brakensiek \& Ragozzine}
\usepackage{natbib}

\newcommand{\be}{\begin{equation}}
\newcommand{\ee}{\end{equation}}
\newcommand{\vt}{\textbf}

\begin{document} 
\newtheorem{thm}{Theorem}

\title{ Efficient Geometric Probabilities of Multi-Transiting Exoplanetary Systems from CORBITS}

\author{Joshua Brakensiek\altaffilmark{1,2} \& Darin Ragozzine\altaffilmark{3,4}}

\email{ jbrakens@andrew.cmu.edu, dragozzine@fit.edu }
\altaffiltext{1}{Homeschool, Phoenix, Arizona}
\altaffiltext{2}{Carnegie Mellon University, Department of Mathematical Sciences, Wean Hall 6113, Pittsburgh, PA 15213.}
\altaffiltext{3}{University of Florida, Department of Astronomy, 211 Bryant Space Science Center, Gainesville, FL 32611}
\altaffiltext{4}{Florida Institute of Technology, Department of Physics and Space Sciences, 150 West University Blvd., Melbourne, FL 32940}


\begin{abstract}
NASA's \emph{Kepler Space Telescope} has successfully discovered thousands of
exoplanet candidates using the transit method, including hundreds of stars with multiple transiting planets. In order to estimate the frequency of these valuable systems, it is essential to account for the unique geometric probabilities of detecting multiple transiting extrasolar planets around the same parent star. In order to improve on previous studies that used numerical methods, we have constructed an efficient, semi-analytical algorithm called CORBITS which, given a collection of conjectured exoplanets orbiting a star, computes the probability that any particular group of exoplanets can be observed to transit. The algorithm applies theorems of elementary differential geometry to compute the areas bounded by circular curves on the surface of a sphere (see Ragozzine \& Holman 2010). The implemented algorithm is more accurate and orders of magnitude faster than previous algorithms, based on comparisons with Monte Carlo simulations. We use CORBITS to show that the present solar system would only show a maximum of 3 transiting planets, but that this varies over time due to dynamical evolution. We also used CORBITS to geometrically debias the period ratio and mutual Hill sphere distributions of \emph{Kepler}'s multi-transiting planet candidates, which results in shifting these distributions towards slightly larger values.  In an Appendix, we present additional semi-analytical methods for determining the frequency of exoplanet mutual events, i.e., the geometric probability that two planets will transit each other (Planet-Planet Occultation, relevant to transiting circumbinary planets) and the probability that this transit occurs simultaneously as they transit their star. The CORBITS algorithms and several worked examples are publicly available at https://github.com/jbrakensiek/CORBITS. 
\end{abstract}

\keywords{celestial mechanics, methods: analytical, methods: numerical, methods: statistical, occultations, planets and satellites: detection, planetary systems}

\clearpage


\section{Introduction} \label{sec:intro}

In the last couple of decades, astronomers have discovered thousands of exoplanets outside of our Solar System through a variety of techniques.  The transit method discovers exoplanets by detecting periodic dimming in the amount of light coming from a star caused by the planet blocking a portion of the star.  NASA's \emph{Kepler Space Telescope} (hereafter "Kepler") has used this method to detect thousands of planet candidates -- a huge step forward -- due to its revolutionary dataset that is far more extensive in duration ($\sim$4 years), duty cycle ($\sim$90\%), and precision than any previous search \citep[e.g.,][]{2015arXiv150604175B}. Due to its ability to detect small planets, Kepler has uncovered a rich vein of Systems with Tightly-packed Inner Planets (STIPs) which often exhibit multiple transiting planets. Observations of systems with multiple transiting planets are extremely valuable for understanding the formation and evolution of this new population of STIPs as well as other planetary systems like our own \citep[][hereafter RH10]{RH10}. One example of the value of these ``multi-transiting systems'' is that they are far more robust to false positives, which has allowed for a joint validation of 851 planets in 340 systems, nearly doubling the number of known exoplanets \citep{2014ApJ...784...45R, 2014ApJ...784...44L}. 

Despite its success, one clear drawback of the transit method is that the planetary orbit must line up with the star from our perspective, otherwise the existence of the planet is near-impossible to infer. Detecting all of the planets in multi-planet systems then becomes very unlikely due to natural deviations from coplanarity, even when mutual inclinations are $\lesssim$1$^{\circ}$. For example, we show below that if an exact replica of the present Solar System were being observed from any direction, at most 3 of the 8 planets would be simultaneously aligned sufficiently well to transit ($\S$ \ref{subsec:appl-ss}). Thus, multi-planet systems extend and complicate the known geometric bias inherent to the transit method. 

We have developed a highly-efficient semi-analytical algorithm for the calculation of multi-transiting geometry which we call CORBITS (Computed Occurrence of Revolving Bodies for the Investigation of Transiting Systems). CORBITS is a practical and critical tool for calculating the geometric bias of multi-transiting systems which we make freely available to the community. After motivating the need for understanding multi-transiting geometry to determine the frequency of planetary systems ($\S$\ref{sec:motivation}, we describe the details of the CORBITS algorithm ($\S$\ref{sec:MTS}). We demonstrate the value of CORBITS by applying it to example questions related to the multi-transiting probability of the solar system and known Kepler multi-transiting systems ($\S$\ref{sec:appl}). After some concluding remarks ($\S$\ref{sec:concl}), we present in an appendix some information on the geometry of exoplanet mutual events, some of which is relevant to transiting circumbinary planets (Appendix A).

\section{Motivation}
\label{sec:motivation}
When planets are considered independently, the geometric detection bias can be computed on a planet-by-planet basis. However, for any exoplanetary properties that depend on correlations between planets, the full multi-transiting geometry must be understood. Extracting information-rich insights from multi-transiting systems therefore requires an efficient method for determining the geometric bias of multi-planet systems (RH10). 

To motivate the importance of understanding multi-transiting geometry, we emphasize that a careful and critical distinction must be made in distinguishing the average Number of Planets Per Star (NPPS) and the Fraction of Stars with Planets (FSWP), following the notation of \citet{2011ApJ...742...38Y}. 

Most frequency/occurrence studies are actually calculating the average Number of Planets Per Star (NPPS). When occurrence is calculated planet-by-planet without accounting for the complex correlations in detecting planets in multiple systems, the result is the average NPPS. Understanding NPPS gives us huge insight into planet formation (e.g. period/radius distribution) and other important questions (e.g. the frequency of potentially habitable planets).

Determining the Fraction of Stars With Planetary systems (FSWP) requires a detailed analysis that self-consistently includes deriving the multiplicity and inclination distributions, as discussed below. In this case, the frequency/occurrence is calculated system-by-system. FSWP is critical for scientific (What is the efficiency of planet formation?) and observational (How should I design my exoplanet survey mission?) questions.

The true multiplicity (the number of planets per star in the actual underlying population) is, by definition, the ratio of NPPS/FSWP. NPPS can be accurately calculated in the present of multi-transiting systems only insofar as all planets in a system are included (in both the search and the subsequent analysis). 

Analyses that debias on a planet-by-planet basis cannot directly estimate FSWP or the average multiplicity even though they can determine NPPS. Using only the most probable planet or the most detectable planet does not accurately handle correlations between transiting probabilities in real exoplanetary multiple systems. The approximation of calculating FSWP by using only the most probable or detectable planet is only correct if all exoplanetary systems are exactly coplanar, which is not supported by the \emph{Kepler} data. Because the transit method is sensitive to inclinations at the $\lesssim$1 degree level, even these small inclinations cause significant errors in the estimation of FSWP that uses only the most probable/detectable planet.

The approximation of using the more probable planet to estimate FSWP infers the number of ``missed planets'' due to geometric debiasing and assigns these to stars that have no planets, but in the non-coplanar case, some of these ``missed planets'' could be in systems with other known planets. Both numerical and analytical models confirm the result that planet-by-planet debiasing techniques will underestimate FSWP by an unknown amount, though generally less than a factor of 2.

Kepler’s multis encode information on the true multiplicity and inclination distributions \citep[e.g.,][]{2011ApJS..197....8L,2012AJ....143...94T,2012A&A...541A.139F,2012ApJ...761...92F,2012ApJ...758...39J,2012arXiv1203.6072W,2014arXiv1410.4192B}. CORBITS allows for accurate calculations of the multi-transiting geometry, thereby enabling methods that can solve for the average multiplicity, the inclination distribution, NPPS, FSWP, and other interesting and important information pertinent to the architectures of planetary systems.


\section{CORBITS Algorithm for Multi-Transiting Geometry} \label{sec:MTS}


Our goal is to understand how the geometric probability of transiting in a system with multiple transiting planets affects our interpretation of multi-transiting systems. Although there are many aspects to this interpretation, our focus here is on the purely geometric component of calculating the bias towards detecting or not detecting planets in multiple systems. With this in mind, we convert an astronomical question into a geometry problem. 

Formally, the problem at hand is this: given a (conjectured) exoplanetary system with all the orbital properties of $N$ planets known, what is the probability that a random observer would be able to observe the transit of a particular $m$-planet subset? This problem can be divided into two parts.
\begin{enumerate}
\item For a system of planets on fixed orbits that is observed perfectly for an infinite time, what is the probability that a specific subset of $m$ out of $N$ planets transit and the other $N-m$ planets do not transit? 
\item How is this idealized case affected under actual observational limitations and for realistic planetary systems? 
\end{enumerate}

Our algorithm addresses the first question, an idealized geometric problem that is very relevant to Kepler multi-transiting systems. This calculation is critical, but the ideal case may not constitute a complete description of the probability of observing real planet transits, primarily for three reasons: duty cycle, signal-to-noise ratio (SNR), and non-Keplerian orbits. These caveats are described in greater detail in $\S$\ref{subsec:MTS-caveats} below.  

We begin with a focus on the ideal geometric question: given $N$ planets with known (fixed) orbital elements, what is the probability that a specific subset of these planets have orbits that cross the parent star from the viewpoint of an isotropically-random observer? For decades, the solution of $\frac{R_*}{a}$ has been known for the probability of detecting transits of one circular planet \citep[e.g.,][]{1984Icar...58..121B}. The eccentric one-planet case modifies this solution to $\frac{R_*}{r}$ where $r=\frac{a(1-e^2)}{1 + e \cos f}$ \citep[e.g.,][]{2010exop.book...15M} is the actual star-planet distance along the line-of-sight. We use the standard variables for orbital elements throughout: $a$ is the planetary semi-major axis, $e$ is the eccentricity, $f$ is the true anomaly, $\omega$ is the argument of periapse, $i$ is the inclination, and $\Omega$ is the longitude of the ascending node. We follow the geometric convention of RH10 and \citet{2010exop.book..217F}, but since we are considering an isotropic observer-averaged probability, the exact coordinate system does not matter.

Before the prevalence of multi-transiting systems, a few quick-but-inaccurate methods were used to determine the probability of the $N \ge 2$ case. To our knowledge, motivated by the discovery of the first multi-transiting systems \citep{2010ApJ...725.1226S}, the first accurate assessment was given by RH10. RH10 relied primarily on a numerical Monte Carlo method, but included an analytic estimate for the two-planet circular case. \citet{2012AJ....143...94T} presented the first semi-analytical solution using, effectively, a spherical harmonic expansion. Their method focused on multiple circular planets with inclinations drawn from a single distribution, which is reasonable for some population studies. 

We present in this work an efficient, accurate, and precise analytical solution to this geometric problem. For simplicity, we will refer to the algorithm as CORBITS, short for the Computed Occurrence of Revolving Bodies for the Investigation of Transiting Systems. The C++ code for the implementation of CORBITS is publicly available at \texttt{https://github.com/jbrakensiek/CORBITS} to encourage the accurate study of multi-transiting systems by the community.  Currently, CORBITS consists of a command-line utility able to be integrated with existing Unix pipelines and a C++ API so that more advanced features of CORBITS can be utilized. We have striven to make this code easy for the community to use. The implementations for most of the examples in Section \ref{sec:appl} are provided on GitHub.

\subsection{Terminology and Methods} \label{subsec:MTS-term}

Our strategy follows \citet{1984Icar...58..121B} and RH10 by considering the entire ``celestial sphere'' of observers around a planetary system. As an exoplanet moves around its star, it casts a shadow on different locations on the celestial sphere as shown in Figure 1.  The union of all these locations is a band centered about a great circle of the sphere.  We shall call this band a \emph{transit region}. The probability of observing an exoplanet transit is now the surface area of the transit region divided by surface area of the sphere.

When the system has multiple planets, these regions become much more complex (RH10).  Figure \ref{fig:geo} shows an example in the three-planet case. (Note that this hypothetical planetary system is shown for illustrative purposes only: the small, nearly-equal, highly-inclined orbits would be completely unstable for real planets.) The calculation of multi-transiting geometry comes down to the determination of the surface area of any arbitrary intersection of these transit regions. 

For simplicity, we begin by describing the CORBITS algorithm in the case of circular orbits($\S$ \ref{subsec:MTS-circle}). In this case, the transit regions on the celestial sphere have a simpler structure which provides intuition for handling the general elliptical orbits ($\S$ \ref{subsec:MTS-ellipse}). 

The transit regions for circular orbits are bounded by two circles, embedded on parallel planes, called \emph{small circles}. If the circular orbits' intersection with the celestial sphere were the ``equator'', these small circles would be latitude lines at North and South $\frac{R_*}{a}$ declination/latitude. Exploiting the properties of small circles makes computations simpler in this case, but it turns out that the general elliptical orbit case can make use of a very similar algorithm. 

The union of regions whose area needs to be computed is bounded by these small circles and their intersections. Practically all techniques of spherical geometry and trigonometry work with great circles only and are not applicable to this problem. For example, these intersection regions are not spherical polygons since the bounding arcs are from small circles.  In the scarce references dealing with finding areas of polygons on the sphere bounded by small circles, these regions are sometimes called ``semi-spherical polygons'' and we will follow this terminology.

\subsection{The Gauss-Bonnet Theorem and Geodesic Curvature} \label{subsec:MTS-geo}

\begin{figure}
\begin{center}
\includegraphics[width = 2in]{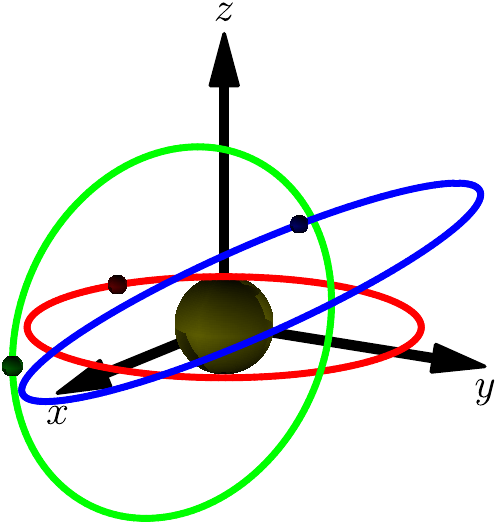}
\includegraphics[width = 2in]{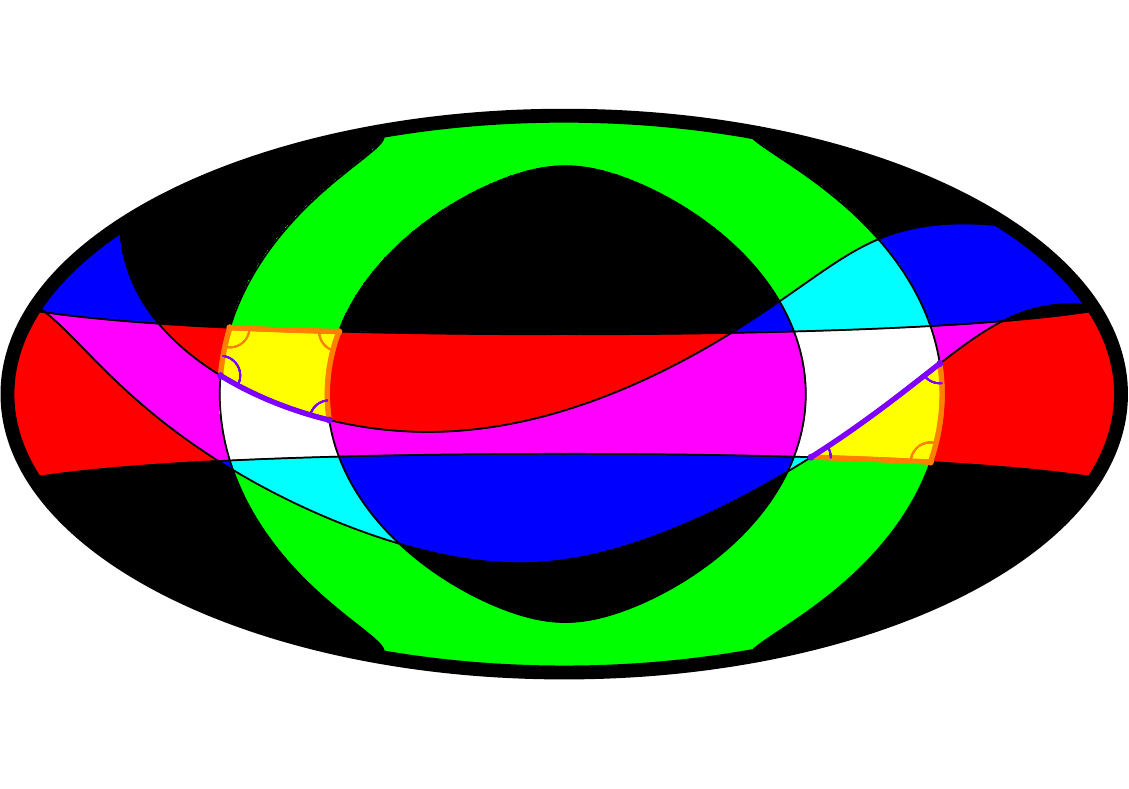}
\end{center}

\caption{Image of a hypothetical three planet system and its corresponding celestial sphere reprojected onto an ellipse.  The blue region is the region of the celestial sphere where distant observers could see only the blue planet transit and similarly for the red and green planets. All the regions are colored based on which planets would appear transiting to distant observers, e.g., the yellow region can see the red and green planets transit as yellow = red + green.  The black region can observe no planets while the white region can observe all of the planets. All of the regions are bounded by ``small circles'' equivalent to latitude lines. The orange and purple arcs show the arc lengths and the turning angles which must be computed in order to calculate the area of the yellow region using the Gauss-Bonnet Theorem as described in the text.  The orange arcs are positive while the purple ones are negative; the signs define which side of the boundary contains the transit.}
\label{fig:geo}
\end{figure}

In order to facilitate the computation of the area of the transit region, we simplify the problem by making use of a theorem of differential geometry, the Gauss-Bonnet Theorem, which converts the two-dimensional area to a one-dimensional integral over the boundary of the region.  In the context of the unit sphere, let $M$ be a region on the surface of a unit sphere bounded by circular arcs and let $\partial M$ be the boundary of $M$.  Then, the surface area of $M$ is
$$4\pi - \int_{\partial M}k_g,$$
where $k_g$ is the \emph{geodesic curvature} of the boundary of $\partial M$ \citep{UGA}.\footnote{Note that the general form of the Gauss-Bonnet Theorem applies to any surface by writing $2 \pi \chi(M)$ instead of $4\pi$, where $\chi(M)$ is the Euler characteristic of the surface. We have substituted $\chi(M) = 2$ as the Euler characteristic of sphere.}

The geodesic curvature is a measure of the curvature of the boundary relative to the curvature of the sphere.  Because the arcs on the boundary of the regions we are considering are small circles, there is a simple formula for the geodesic curvature in this case \citep{UGA}.  In the case of a circular orbit, for example, $$k_g = \frac{R_*}{\sqrt{a^2 - R_*^2}}.$$
To compute the geodesic curvature of the boundary of a semi-spherical polygon (like our transit regions), it is sufficient to compute the arc lengths of each of the ``edges'' and then weight them by each one's geodesic curvature. To correct for the discontinuities at the intersection points of the arcs, one must add the angles between tangent vectors at each intersection point, called \emph{turning angles} \citep{UGA}. 

Another detail to note is that each component of the geodesic curvature, including the turning angles, can either be positive or negative. The sign of each arc specifies which side of the arc contains the desired transit region and the sign of the turning angles is the product of the signs of the two adjacent arcs, as illustrated in Figure \ref{fig:geo}.

\subsection{Illustration: Two-planet case} \label{subsec:MTS-2}

    For the sake of exposition of the application of the Gauss-Bonnet theorem to this problem, we start by looking more closely at the two-planet circular-orbit case. Up to symmetry, every such system can be defined by four parameters: the semi-major axes $a_1$ and $a_2$ of the two planets, the radius $R_*$ of the star, and the angle $\phi$ of true mutual inclination between them (RH10). 
    
    In fact, we can reduce these four parameters to only three.  If we assume that the radius of our celestial sphere is $1$, then distance between the two parallel planes defining each planet's transiting region is $2R_*/a_1$ and $2R_*/a_2$. Thus, we define the half-widths $h_1 = R_*/a_1$ and $h_2 = R_*/a_2$; these are equivalent to the transit probabilities of each individual planet. Thus, $h_1$, $h_2$, and the mutual inclination $\phi$ are sufficient to calculate the probability of observing both planets in transit, the probability of observing only planet 1 or 2, or the probability of detecting no transits.
    
    Throughout our discussion, we use $h_j\equiv R_*/a_j$ as the half-thickness of the transit region of planet $j$. This corresponds specifically to the probability of transit with impact parameter less than or equal to exactly 1, i.e., the probability that the center of the planet crosses over any portion of the star. A (low SNR) grazing transit is possible when the impact parameter is less than 1 + $\frac{R_p}{R_*}$, corresponding to a slightly larger region on the sky with half-thickness $h_j=(R_* + R_p)/a$. Similarly, requiring a non-grazing transit where the entire planet passes over the star corresponds to using a half-thickness of $h_j=(R_*-R_p)/a$. In practice, these nuances on the transit probability can be controlled by the CORBITS code which allows each planet to have its own limiting impact parameter, allowing for any of these three cases or a user-defined case. With this in mind, we will not distinguish between these cases in our further discussion. 
    
    In the two planet case, there are distinct topologies of the transit regions based on the value of $\phi$ (RH10).  These three cases are, in order of increasing $\phi$: Case 1) one transit region is completely contained within the other transit region; Case 2) neither transit region is completely contained within the other, but the intersection of the two regions is still a complete band; and Case 3) the intersection of the two regions is two disjoint semi-spherical quadrilaterals (which corresponds to looking along the lines of nodes of the planets). There is always a non-zero probability of observing both planets in transit. These three cases can be differentiated analytically using succinct formulas.  Consider the following two values:
\begin{align}    
\mu_1&=\frac{h_1^2+h_2^2-2h_1h_2\cos\phi}{\sin^2\phi}\\
\mu_2&=\frac{h_1^2+h_2^2+2h_1h_2\cos\phi}{\sin^2\phi}
\end{align}
These values define the boundaries between the three cases: Case 1 occurs if and only if $\mu_1 \ge 1$ and $\mu_2 > 1$; Case 2 occurs if and only if $\mu_1 < 1$ and $\mu_2 > 1$; and Case 3 occurs if and only if $\mu_1 < 1$ and $\mu_2 \le 1$.

    RH10 gave an analytical approximation for the transit probabilities in these cases.  In Case 1, the probability is exactly \begin{align}
\min\left(\frac{R_*}{a_1}, \frac{R_*}{a_2}\right) &= \min(h_1, h_2). \label{eq:case1}
\end{align}
In Cases 2 and 3, although they did not distinguish between these cases, they proposed the estimate
\begin{align}
\frac{R_*^2}{a_1a_2\sin(\phi)} &= \frac{h_1h_2}{\sin(\phi)}. \label{eq:case23}
\end{align}
    
With CORBITS, described fully below, we compared these analytical approximations to the exact probabilities of known Kepler Objects of Interest. As discussed in $\S$\ref{subsec:appl-KOI}, the RH10 results are valid to first order, though for Case 2 are often off by a factor of 2 or more. We also describe in $\S$\ref{subsec:appl-KOI} the fraction of KOIs that fall into the three different cases: for the real \emph{Kepler} data and a reasonable inclination distribution, about half of the planets fall within Case 2 or 3 and therefore need a detailed geometric interpretation. 

\subsection{Implementation of $N$ circular planets} \label{subsec:MTS-circle}

An exact analytic solution to even the 2-planet case is difficult to present in closed form and the extension to $N$ circular orbits is much trickier. We now revert to a numerical algorithm to describe the exact (to near machine precision) solution for the case of $N$ circular orbits. This requires calculating the area on the celestial sphere from which observers would see the desired configuration (certain planets transiting and all the rest of the planets not transiting), which we call the ``desired region''. Each desired region is one of $2^N$ probabilities (corresponding to including or excluding each individual planet transit region) which sum to 1. Unlike in the 2-planet case, it is possible that the desired region in the $N\ge3$ planet case does not exist, in which case the desired multi-transiting probability is 0. 

First, the orbits are input into CORBITS.  The user supplies the effective radius $R^{*}$ of the star and the orbital elements $a$, $e$, $\omega$, $\Omega$, and $i$. To supply an limiting impact parameter $b$, the user may input $bR^{*}$ as the effective radius of the star on a planet-by-planet basis.

Second, the orbits are converted into regions on the celestial sphere.  While considering circular orbits, each transit region $t_j$ is described by two parameters:  the half-thickness $h_j$ of the transit region (half the distance between the two circles) and the pole $p_j$ perpendicular to both bounding planes of the transit region.  These parameters are computed directly from the orbital elements.

Third, the intersection points of the boundaries of these transit regions are computed.  Because the boundary of each transit region is defined by the intersection of the celestial sphere with two unique planes, the intersection of the two boundaries is the union of four potential plane-plane-celestial sphere intersections.  With this knowledge, these points can be found with a computational geometry procedure which determines the intersection line of each pair of planes, and the intersection of each of those lines with the celestial sphere.

Fourth, CORBITS identifies the intersection points which are on the boundary of the desired region(s). In the N-planet case, it is often possible for a specific subset of planets to have no common transit regions. In this case, we output a probability of zero and the algorithm terminates. In the non-trivial case, we are left with only the points on the boundary of the intersection of all of the possible desired regions -- there can be one or multiple distinct semi-spherical polygons depending on the geometry.

Fifth, CORBITS determines which arcs are on the boundary of the transit region. We can then compute the geodesic curvature of all the arc lengths using the formula:
$$\displaystyle \sum_j \frac{\sqrt{1 - r_j^2}}{r_j}\theta_j,$$
where $r_j$ is the radius of the small circle, and $\theta_j$ is the sum of the angles subtended by the arcs on the small circle. Since the radius of each small circle is $$r_j = \sqrt{1 - h_j^2}$$ and the number degrees of each arc can be calculated by a trigonometric routine, the geodesic curvature is easily calculated.

Sixth, the turning angles are computed.  Since each intersection point is defined by the intersection of two planes, we can construct at each intersection point two vectors tangent to the celestial sphere, each one parallel to one of the intersecting planes.  The angle between these two vectors is the turning angle.  Due to the convention of the Gauss-Bonnet theorem, we choose the angle modulo $2\pi$ which lies in the range $[-\pi, \pi]$ \citep{UGA}.

Seventh, the total geodesic curvature is calculated.  This is a simple sum after accounting for the sign of each arc and angle.

Eighth, the area of the desired joint transit region is outputted.  With the geodesic curvature computed, by the Gauss-Bonnet theorem the area of the region $M$ is $$4\pi - \int_{\partial M} k_g.$$ We then divide our area for $M$ by $4\pi$, the area of the unit sphere, to get the final probability.

With the help of analytic geometry, CORBITS calculates the area of arbitrarily complex semi-spherical polygons on the celestial sphere which correspond to the probability that a particular subset of $m$ planets transit, while the other $N-m$ do not. 

\subsection{Extension to Elliptical Orbits} \label{subsec:MTS-ellipse}

The extension of this pipeline to elliptical orbits requires modifications to the above algorithm.  In the case of an elliptical orbit, the boundary of its transit region is non-circular, but at all reasonable eccentricities the boundary is well-approximated by non-parallel small circles \citep[see Figure 3 of][]{2010exop.book...55W}.  Such small circles can be chosen so that they are tangent to the true boundary of the transit region while the surface area of the celestial sphere between the two small circles is equal the surface area of the true transit region.  Thus, this approximation is perfect in the one-planet case, and has high accuracy in multi-planet cases.  

The probability of transit of a single elliptical planet is
$$\left(\frac{R}{a}\right)\left(\frac{1}{1-e_i^2}\right).$$
Thus, in order to preserve the exact probability in the one-planet case, we let $$h_i = \left(\frac{R}{a}\right)\left(\frac{1}{1-e_i^2}\right).$$
Let $\vt v_i$ be the unit vector normal to the orbital plane and $\vt w_i$ be the unit vector in the direction of periapse.  Let $$\psi = \pi - \sin^{-1}\left(h_i * (1 - e)\right) - \cos^{-1}(h_i).$$  The vectors orthogonal to the two planes bounding the approximated transit region are then
\begin{align}
\vt p_{i1} &=\vt v_i\sin (\psi) + \vt w_i\cos (\psi),\\
\vt p_{i2} &=- \vt v_i\sin (\psi) + \vt w_i\cos (\psi).
\end{align}
The planes orthogonal to $\vt p_{i1}$ and $\vt p_{i2}$ at distances of $h_i$ from the center of the celestial sphere bound a region of the same area as the true transit region and are tangent to the boundary of the transit region.  Note in particular that when $e=0$, $\psi = \pi / 2$; therefore $$\vt p_{i1} = -\vt p_{i2} = \vt v_i.$$ 

The modifications to the pipeline to account for elliptical planets are as follows. In the first step, the user supplies the parameters $a$, $e$, $i$, $\Omega$, $\omega$, and $bR^*$ for each planet.  In the second step, we construct the approximated transit region using the process described to determine $\vt v$ and $\vt w$.  As the third through eighth steps of the pipeline deal the with abstraction of regions on the celestial sphere which are bounded by small circles, a similar implementation is employed. Since this method supersedes the method for circular orbits, only the elliptical version is available in the public version of CORBITS. For exoplanetary systems, the errors due to our approximations for eccentric orbits are small enough to be practically meaningless, as discussed below. 

\subsection{Comparison to Monte Carlo} \label{subsec:MTS-MC}

We have tested CORBITS by comparing the solutions to a Monte Carlo estimate of multi-transiting probability as in RH10. In so doing, we reproduced the results of RH10 and actually discovered a minor error in the legend of their Figure 2. 

As further discussed in $\S$\ref{subsec:appl-kep11}, we also reproduced Figure 4 from \citet{2011ApJS..197....8L}, which was created using a separate Monte Carlo code. 

Due to its analytical nature, CORBITS is very fast, requiring much less than a CPU-millisecond even for complicated planetary systems. It is also accurate to near double floating-point machine-precision; in practical use, errors in stellar sizes or planetary orbital parameters are much more important than errors induced by our calculations. Accurate probabilities can be important for correctly understanding unusual systems discovered by \emph{Kepler} or the upcoming \emph{TESS} \citep{2014SPIE.9143E..20R} and \emph{PLATO} \citep{2014ExA....38..249R} missions, where probabilities of $\sim$10$^{-7}$ are sometimes useful to evaluate. Random Monte Carlo methods would require a near-intractable number of calculations to reach the precision and accuracy of our model. In fact, CORBITS can determine whether the probabilities are equal to zero, which is technically impossible for simple Monte Carlo methods to determine.

Note that the eccentric case is not an exact solution and the approximation worsens as eccentricity increases and periapse decreases. For small eccentricities $(e \lesssim 0.2$) expected from most multi-transiting systems \citep[e.g.,][]{2011ApJS..197....1M,2014ApJ...790..146F}, the accuracy of the multi-transiting probability is dominated completely by unknown orbital and stellar parameters and not by the approximation in CORBITS. To investigate a worst-case scenario, we examined an extreme 2-planet system composed of the one of the most eccentric transiting planets, HD 80606b \citep[e=.93366][]{2009arXiv0906.4904G} and a hypothetical interior planet with a periapse at the stellar Roche limit and apoapse at the periapse of planet b. This is a worst-case scenario that maximizes the inaccuracies in our technique. We ran CORBITS and a Monte Carlo comparison for a large range of random orientations. The typical fractional error in the multi-transiting probability from CORBITS was $\sim$0.3\% and the worst-case fractional error was $\sim$3\%. Even in this most extreme imaginable case, the errors are likely dominated by errors in stellar parameters (often known to only $\sim$5-10\%). We are satisfied that the eccentric approximation of CORBITS is sufficient for practical usage.

\subsection{Caveats of CORBITS} \label{subsec:MTS-caveats}

Although the Monte Carlo simulations establish that CORBITS has a high degree of accuracy, there are some limitations on the scope of CORBITS. Since the planet has a random orbital phase and since no dataset contains 100\% monitoring (even Kepler only had a $\sim$90\% duty cycle), there is always a chance that the transit of a planet occurred, but was missed in the observations. There is an additional ``geometric effect'' when the planet's orbital period is longer than the observing baseline; this factor is simply (baseline/period) for a single planet. For multiple long-period planets, the effect depends on the specific positions of the planets and the times of their observations in a way that would need to be modeled separately. If multiple transits are required for identifying a planet, this must be accounted for as it may significantly reduce the probability of detecting planets with periods near the observational baseline; practically speaking, this requires calculation of the window function for the observations of interest \citep[e.g.,][]{2014ApJ...792...79B}. 

Even when a transit is present, it may not be detectable due to insufficient signal-to-noise ratio. Calculating the true effective "Signal" and "Noise" for a transiting planet can be complicated, as can be estimating whether a specific SNR would actually be detected with any specific transit-search pipeline. Completeness studies \citep[e.g.,][]{2013ApJS..207...35C,2013ApJ...766...81F,2013ApJ...770...69P,2015arXiv150604175B} address these issues. We ignore this aspect of detection probability as non-geometric. 

Finally, our model assumes fixed Keplerian orbits. It is well known that planet-planet interactions violate this assumption. Such interactions are observed as transit timing variations (TTV) as predicted by \citet{2005MNRAS.359..567A} and \citet{2005Sci...307.1288H} and seen abundantly in the Kepler data \citep[e.g.,][]{2013ApJS..208...16M}. These TTV signals are due almost entirely to changes in planetary phase along the orbit, which is not important for computing the geometric transiting probability. Actual modifications to the orbits themselves on observational timescales are almost entirely negligible in the case of multiple planets; this can be seen empirically by the fact that no known \emph{Kepler} planet candidates have come into or out of transit. In fact, even slight changes in planetary orbits that would manifest themselves as slow changes in the transit shape are very rare \citep[][Fabrycky et al. 2015, in prep., Becker \& Adams, in prep.]{2002ApJ...564.1019M, 2009ApJ...698.1778R}. Therefore, the Keplerian approximation is a very good one for short timescales. 

One might consider that the true random probability of observing specific planets in transit would need to be averaged over long dynamical times since eccentricities, apsidal angles, inclinations, and nodal angles all vary as a function of time \citep[][Becker \& Adams, in prep.]{2000ssd..book.....M}. Our observations come at a random time in the dynamical evolution for these systems. The easiest way to calculate these probabilities would be to use CORBITS on snapshots of the system taken from an analytical (i.e., Laplace-Lagrange) or numerical (i.e., from an N-body integration) analysis and to combine these into a final probability. We perform an example of this analysis for the solar system as one of our applications ($\S$\ref{subsec:appl-sssecular}). While this is an interesting and potentially important problem, a full solution is beyond the scope of the present paper and we leave it for future work.


\section{Applications} \label{sec:appl}

We anticipate that CORBITS will be used by the exoplanet community for a wide variety of problems relating to multi-transiting systems. To illustrate the importance of correctly accounting for multi-transiting probabilities and the value of the CORBITS algorithm, we perform some investigations here. These include application of CORBITS to the present-day solar system ($\S$\ref{subsec:appl-ss}) and its long-term evolution ($\S$\ref{subsec:appl-sssecular}). We use CORBITS to debias the period ratio distribution of \emph{Kepler} multi-transiting systems ($\S$\ref{subsec:appl-per}) followed by a similar analysis for the mutual Hill sphere distribution ($\S$\ref{subsec:appl-mhs}). As mentioned above, we benchmark our code by reproducing the expected inclination distribution for Kepler-11 and Kepler-90 ($\S$\ref{subsec:appl-kep11}). We then look at how multi-transiting geometry affects the entire \emph{Kepler} candidate list ($\S$\ref{subsec:appl-KOI}) Most of these examples are available on the CORBITS GitHub repository so that the exoplanetary community can directly investigate and/or adapt these examples.

\subsection{Applications of CORBITS to the Solar System} \label{subsec:appl-ss}

To demonstrate the utility of the implemented pipeline, we performed simulations of the Solar System from the perspective of a random distant observer. We used the 8 planetary orbits at epoch J2000 from JPL HORIZONS. Interestingly, we find that a distant observer could see at most three distinct transiting planets (assuming sufficiently long observations and fixed Keplerian orbits, as discussed in Section \ref{subsec:MTS-caveats}). Specific results for the solar system can be seen in Table \ref{tbl:ss-time}.  In particular, the table shows all triples of planets whose probability is non-zero.

This result clearly demonstrates the highly-biased nature of transit observations in exoplanetary systems. In particular, as larger planetary orbits are included, the mutual inclination between planets required to see them all transit shrinks significantly below the $\sim 1^{\circ}$ dispersion seen in the Solar System and exoplanetary systems \citep[e.g.,][]{2011ApJS..197....8L,2012ApJ...761...92F,2014arXiv1410.4192B,2014ApJ...790..146F}. It also highlights the value of using radial velocities and transit timing (or duration) variations to infer the presence of non-transiting planets \citep[e.g.,][]{2012Sci...336.1133N,2014ApJS..210...20M}

%

\begin{deluxetable}{lll}
\tablecaption{Solar System Transit Probabilities}
\tablehead{\colhead{Planets} & \colhead{Probability} & \colhead{$N$}}
\startdata
Mercury-Venus & $6.84 \times 10^{-4}$ & $1460$\\
Earth-Venus & $3.22\times 10^{-4}$ & $3100$ \\
Earth-Mars    & $2.84 \times 10^{-4}$  & $3520$\\\hline
Mercury-Earth-Mars & $2.10 \times 10^{-4}$ & $4750$\\
Mercury-Venus-Saturn & $3.53 \times 10^{-5}$ & $28300$\\
Mercury-Earth-Uranus & $6.15\times 10^{-8}$ & $1270000$\\
Venus-Earth-Uranus & $2.17 \times 10^{-5}$ & $46100$\\
Mercury-Mars-Uranus & $4.95 \times 10^{-6}$ & $202000$\\
Jupiter-Saturn-Uranus & $2.22 \times 10^{-6}$ & $450000$\\
Mercury-Venus-Neptune & $1.77 \times 10^{-6}$ & $565000$\\
Mars-Jupiter-Neptune & $2.56 \times 10^{-6}$ & $391000$\\\hline
Exactly 2 Planets & $2.10 \times 10^{-3}$ & $476$\\
Exactly 3 Planets & $2.79 \times 10^{-4}$ & $3580$\\
At least 4 Planets & $0$ & $-$\\
Inner Planets & $0$ & $-$\\
Outer Planets & $0$ & $-$\\\hline
Kepler 11 b-c-d-e-f-g & $1.80\times 10^{-4}$ & $5550$\\\hline
\enddata
\tablecomments{The probability of multiple solar system planets transiting as determined by the CORBITS algorithm. The first column shows example sets of planets. The second column lists the probability of observing all the planets in the set transit, assuming the present-day orbits and long-term observations. A few pairs of interest and all non-zero three-planet probabilities are listed. In the case of Solar System planets, the probabilities assume that the observer is at a random location outside our Solar System.  The third column lists the inverse of the probability, i.e., the expected number of randomly-oriented analogs of this planetary system needed to expect to find $\sim$1 system where this subset transits. For comparison, the probability of observing all 6 planets in the Kepler-11 system (assuming a typical inclination distribution) is also shown ($\S$\ref{subsec:appl-kep11}).\label{tbl:ss-time}}
\end{deluxetable}

\subsection{Time Evolution of Solar System Multi-Transiting Geometry}

\label{subsec:appl-sssecular}
As discussed in the CORBITS caveats ($\S$\ref{subsec:MTS-caveats}), the probabilities computed in the previous section on the present-day solar system will change as the orbits of the planets evolve due to gravitational interactions. Nevertheless, the subsequent evolution of the multi-transiting probability can be readily explored using CORBITS. 

We begin with the secular approximation of the evolution of the solar system from \citet{2014ApJ...790..146F} which starts at the present epoch and evolves the system for 5 MYr. This is not exactly the same as the evolution represented by a full n-body integration, but is sufficiently accurate for our purposes of illustrating the multi-transiting geometry. Snapshots are taken of the solar system's instantaneous osculating orbital elements every 0.01 MYr and sent to CORBITS, which computes the multi-transiting probability for all 64 possible cases of including and excluding each of the 8 planets in transit. 
We find the behavior to be generally as expected: as planetary inclinations and nodes vary, the probability for any pair of planets to both transit can change substantially, with strong enhancements when nodes are nearly aligned. It is important to remember that, due to the combination of forced and free inclinations and nodes, it is not simply the case that each planet circulates through all possible node values \citep{2000ssd..book.....M}; this secular solution captures that aspect of real systems.

\begin{figure}
\begin{center}
\includegraphics[width = 3in]{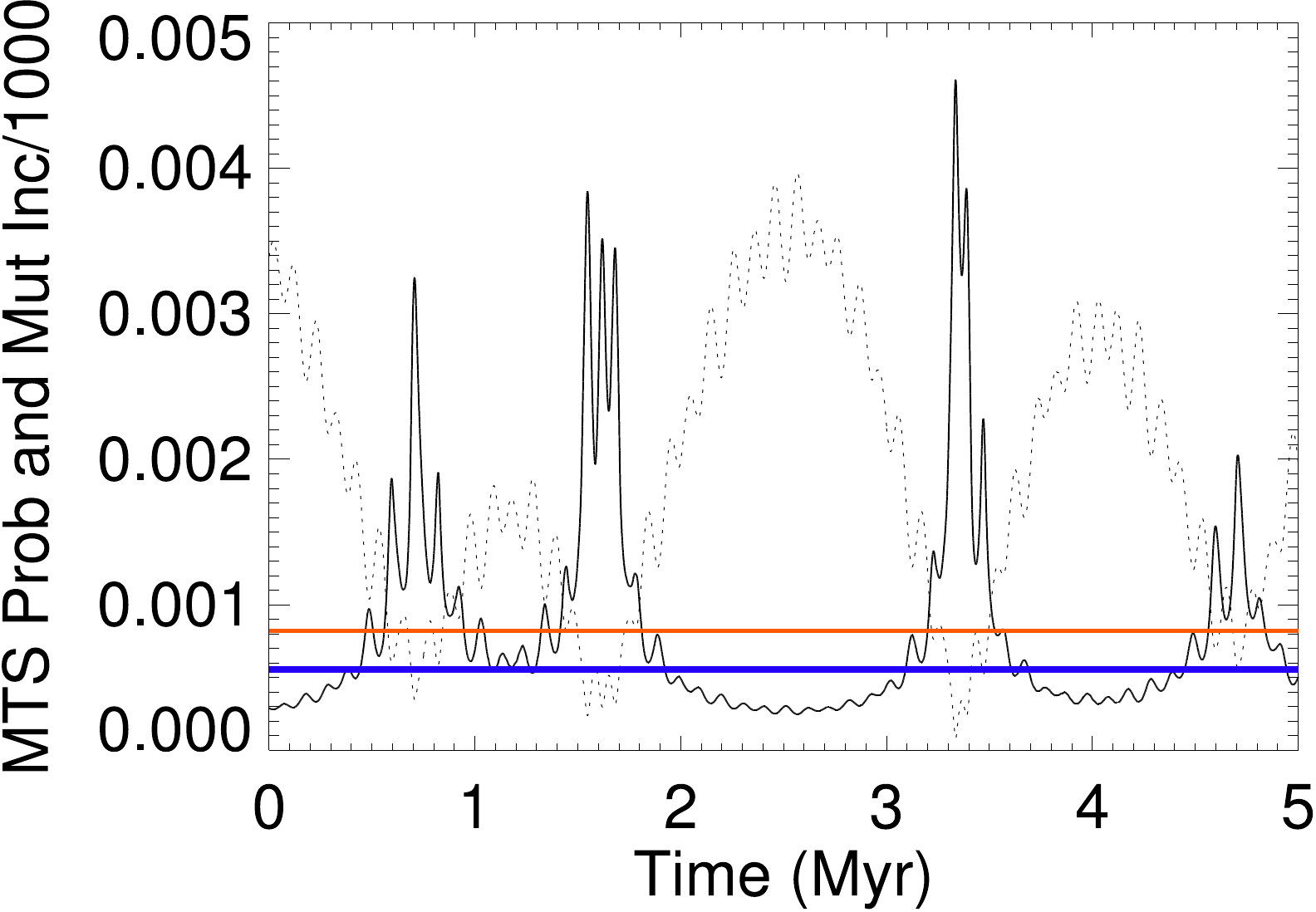}
\end{center}

\caption{Variations in the Multi-Transiting Probability for Venus and Earth due to orbital interactions. The solid line shows the time evolution of the Venus-Earth double-transiting probability evaluated by CORBITS at 0.01 MYr steps in the \citet{2014ApJ...790..146F} secular model of the solar system. The dashed line shows the Venus-Earth mutual inclination in degrees divided by 1000, e.g., the Venus-Earth mutual inclination varies from 0.1$^{\circ}$-4$^{\circ}$. As expected, there is a clear anti-correlation between the mutual inclination of the two orbital planes and the double-transiting probability. This anti-correlation is rather non-linear due to the unique properties of multi-transiting geometry (see discussions of Cases 1-3 and Equation \ref{eq:case23} in $\S$\ref{subsec:MTS-2}). The lower thicker blue line shows the time-averaged probability (the median), while the upper thinner orange line shows the probability-weighted time-averaged probability (the mean). See the text for additional discussion.}
\label{fig:vemtprob}
\end{figure}

Of great interest is the Earth-Venus pair, which shows the expected clear anti-correlation between mutual inclination and double-transiting probability in Figure \ref{fig:vemtprob}. Note that this anti-correlation is quite non-linear due to the different double-transiting regimes (Cases 1-3 from $\S$\ref{subsec:MTS-2}). The Venus-Earth double-transiting probability can spike to be over 15 times the present-day amount, but these high probability spikes are short lived. Therefore, two summary statistics are of interest: the time-averaged probability (given by the median) and the probability-weighted time-averaged probability (given by the average). If Venus and Earth were discovered as a double-transiting system, the probability-weighted time-averaged probability would be the appropriate one to use for debiasing (e.g., estimating the number of Venus-Earth analogs that would be present in the underlying population). 

As seen in Figure \ref{fig:vemtprob}, the probability-weighted double-transiting probability is $\sim$3 times the present day double-transiting probability. However, such a difference would roughly ``average out'' if many systems were observed; this is a key motivation of studying the ensemble of \emph{Kepler} multi-transiting systems, though detailed study would be needed to estimate these probabilities averaged over dynamical timescales. 

Another consequence of the enhanced probability of detecting double-transiting systems when they have lower mutual inclinations is that the present-day inferred inclination distribution must be an underestimate of the inclination distribution that would be inferred if dynamical evolution was considered. However, we note here that Venus-Earth experience secular variations in inclination that are larger than the critical inclination that separate Cases 1-3. Planets in STIPs are much closer to their stars, so similar inclination variations would not have such a large effect. Scaling from the Venus-Earth example, a rough estimate is that dynamical enhancement of multi-transiting probabilities among \emph{Kepler} systems are underestimated by the typical inclination distribution by $\sim$10\% or $\sim$0.2$^{\circ}$, currently a small effect. CORBITS could be profitably used to investigate this effect specifically (see also Becker \& Adams, in prep.). 

Also of interest is the time-evolution of the number of transiting planets. The probability of finding any solar system planet transiting varies from 1.9\% to 2.5\%, with a probability-weighted time-averaged probability of around 2.4\%. This variation is primarily due to the fact that when multiple planets transit, if their transit regions overlap one another then the probability of finding any one planet in transit is diminished. The ratio of the probability of seeing only one planet transit to the probability of seeing multiple planets transit (e.g., single vs. multi) varies from 2-10, with a probability-weighted time-averaged value of about 6. 

\begin{figure}
\begin{center}
\includegraphics[width = 3in]{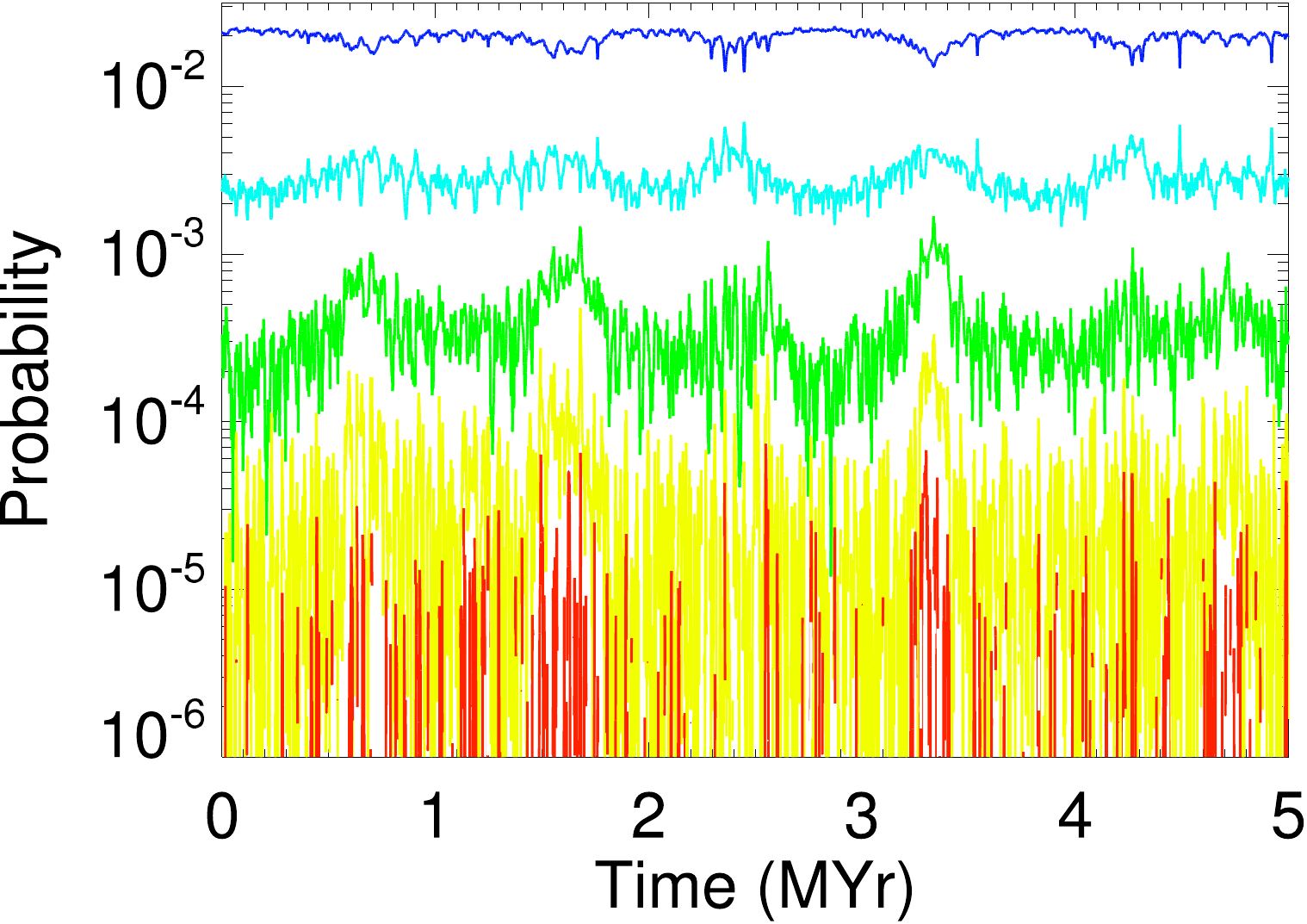}
\end{center}

\caption{The Probability of Observing Different Numbers of Transiting Planets in the Solar System due to Dynamical Evolution. Using the secular evolution of the solar system from \citet{2014ApJ...790..146F}, CORBITS can determine the probability that any subset of solar system planets would be seen to transit by a randomly-oriented distant observer. The apparent multiplicity can be tabulated from the 64 possible combinations of transiting and non-transiting planets. Here we show the dynamical evolution of this apparent multiplicity, with probability shown on a log scale. The different colors indicate, from top to bottom, the probability of seeing exactly 1, 2, 3, 4, and 5 planets in transit. There are rare occasions when 6 planets have a very low probability to transit which are not shown; over the timeframe shown here, more than 6 planets are never seen to transit. As \emph{Kepler} systems are much more compact, this graph is not representative of these systems.}
\label{fig:ssmtprob}
\end{figure}

While only 3 planets in the current solar system could be seen in transit by any outside observer, there are times when up to 6 planets are seen in transit, though with very low probabilities. This result is shown in Figure \ref{fig:ssmtprob}. More seriously, the apparent multiplicity, averaged over all observers, can also vary as the probability of exactly $m$ planets transiting fluctuates.  

We emphasize again that \emph{Kepler} observations are dominated by STIPs which are in a different regime from the solar system that mitigates the effects of dynamical evolution on multi-transiting probabilities. Still, adding dynamical evolution to the determination of multi-transiting probabilities adds significant complications and will require detailed future investigations. 

\subsection{Application to Period Ratio Distribution of Multi-Transiting Systems} \label{subsec:appl-per}

Period ratios in exoplanetary systems are a crucial diagnostic of their formation, evolution, and dynamics. In particular, period ratios that are near a ratio of two small integers are known to exhibit unique and heightened dynamical behavior due to mean-motion resonances. Dissipative dynamics are thought to drive planets into resonant configurations; subsequent planet-planet scattering or stochastic migration forces can randomize the period ratios. Thus, the ratio of \emph{Kepler} orbital periods in exoplanetary systems is a critical test of planet formation theories \citep[e.g.,][]{2012MNRAS.427L..21R,2013MNRAS.436L..25M,2013ApJ...778....7B,2014MNRAS.445..749H,2014MNRAS.440L..11R,2015ApJ...798L..32C,2015ApJ...807...44P}

\emph{Kepler} has provided enormous insight into the exoplanetary period ratio distribution. First, \emph{Kepler} has identified an order of magnitude more planetary systems to date than all other methods and observations combined. Second, possible eccentricity-dependent period aliases in radial velocity observations provide a difficult-to-characterize source of confusion \citep[e.g.,][]{2010ApJ...709..168A}. Finally, with the exception of a very rare aliases, \emph{Kepler} transiting planet candidate periods and period ratios are very precisely measured, an important consideration when identifying resonant or near-resonant behavior. 

\citet{2011ApJS..197....8L} performed the first investigation of the period ratio distribution of \emph{Kepler} candidates. They found that the majority of systems did not show a preference for strong resonances. Still, there was a statistically significant excess of planets just wide of first-order resonances, especially the 3:2 and 2:1. Subsequent studies with more candidates find similar results \citep{2014ApJ...790..146F,2015MNRAS.448.1956S}. 

These studies have focused on the observed period ratio distribution (\citet{2015MNRAS.448.1956S} being an important exception). We improve upon these distributions by accounting for the observational bias from multi-transiting geometry, thus determining the nature of the true underlying period ratio distribution. As pointed out in RH10, when mutual inclinations are larger than the critical inclination, whether two planets are both seen to transit is a modified product of the independent transiting probabilities. Since larger period ratios generally necessitate that one planet be further from the star and thus have a smaller transiting probability, there is some bias against finding planets with large period ratios. Therefore, it is clear that the observed distribution may be biased geometrically towards having smaller period ratios. As discussed in \citet{2013ApJ...763...41C}, there is also a general observational bias against longer periods since the same planet orbiting further away would also have a decreased chance of detection.

We begin by defining a sample of \emph{Kepler} candidates that are mostly free from observational biases. We use as our database of planet candidates all systems which are not classified as "False Positive" in a full download\footnote{Available at \texttt{http://exoplanetarchive.ipac.caltech.edu/cgi-bin/ExoTables\\/nph-exotbls?dataset=q1\_q17\_dr24\_koi}.} of the ``Kepler Objects of Interest - Q1 through Q17 DR24 KOI'' Table as of July 16, 2015. Our sample includes 1353 pairs of candidates in systems with 2 or more planets or planet candidates. The data and codes used to perform these calculations are available in the CORBITS code package. 

These 1353 pairs of candidates do not include KOI-284 and KOI-2248, which we removed since these systems are unlikely to be true multis \citep{2014ApJ...790..146F}. We also remove Kepler-90 (KOI-351) which appears to be an outlier in having an unusually low multi-transit probability for typical mutual inclinations (see $\S$\ref{subsec:appl-kep11}). 

To ensure that we have an observationally complete sample, we employ an SNR cut of 16 and an impact parameter cut of 0.8 as in \citet{2013ApJ...763...41C}. To remove the observational bias against planets at long periods, we follow \citet{2013ApJ...763...41C} and \citet{2011ApJS..197....8L} by taking every pair of planets and recalculating the SNR if their orbital periods were reversed, using SNR $\propto P^{1/3}$. Only if both planets can be detected at each other's periods do we keep the pair; this usually results in removing pairs with inner planets just above the SNR cut. Technically, this is stricter than needed for debiasing just the period ratio distribution (as opposed to the radius ratio distribution studied in  \citet{2013ApJ...763...41C} and \citet{2011ApJS..197....8L}) . It has the effect of showing the period ratio distribution independent of radius.

Note that in systems with high multiplicity, this calculation is performed pair by pair and not system by system. Although there is more physical meaning to the period ratio between neighboring planets, it is generally impossible to know for certain whether two planets are indeed neighboring, since there could be intermediate small or non-transiting planets. This could be accounted for in an extension of the method used in \citet{2011ApJS..197....8L}: a forward model of simulated planetary systems that are then ``observed'' by a simulated \emph{Kepler} could determine the underlying period ratio distribution of neighboring pairs in a much more powerful way. However, expanding that model is beyond the scope of this work and we generally just consider every possible period ratio. 

After these cuts, 556 period pairs remain; these are free from the dominant SNR-related observational biases. We then calculate every period ratio, ${\cal P} \equiv P_{outer}/P_{inner}$, so that ${\cal P} > 1$ and only consider period ratios $1 < \cal{P} < 4$. Beyond a period ratio of 4, there is little meaningful structure in the period ratio distribution. 

We can now use the CORBITS code to remove the geometric observational bias from these systems. As we have seen, this requires choosing a mutual inclination for each pair. Past investigations have identified that typical \emph{Kepler} planets follow an inclination distribution of a Rayleigh distribution with width of $\sim$1.5$^{\circ}$ \citep[e.g.][]{2011ApJS..197....8L,2012ApJ...761...92F,2014arXiv1410.4192B}. We employ a Monte Carlo method that takes each pair and draws $10^4$ Monte Carlo samples with this distribution. In combination with this inclination, we use the nominal parameters in the exoplanetarchive Table, to calculate the average probability that both planets in the pair are transiting and give each value of ${\cal P}$ a weight of the inverse of this probability. Note that this gives more weight to pairs with longer orbital periods, as it should. The weighted distribution is now an approximation to the true underlying distribution since we have removed the major detection and geometric biases. 

\begin{figure}
\begin{center}
\begin{tabular}{cc}
\includegraphics[width=0.23\textwidth]{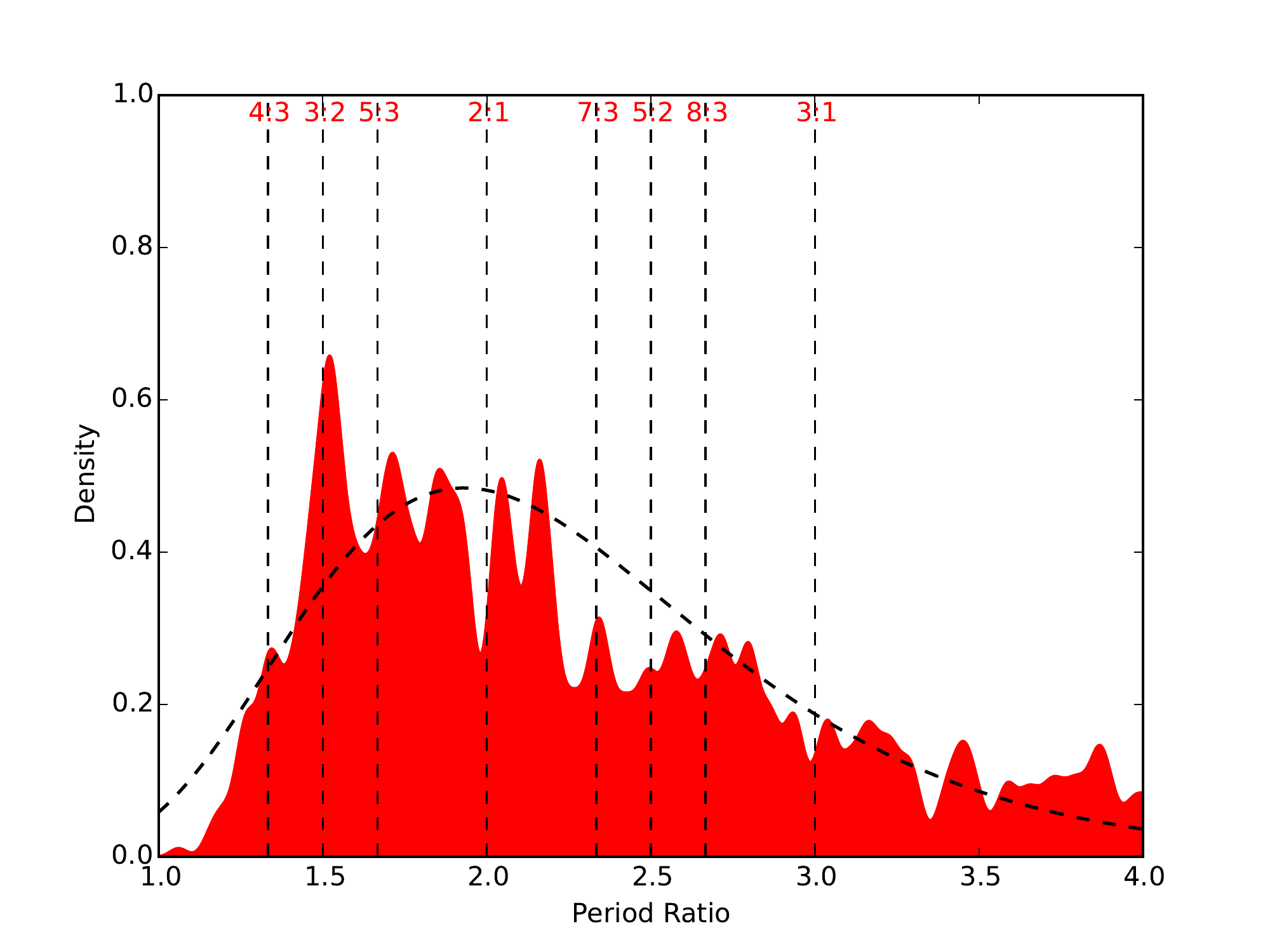} &
\includegraphics[width=0.23\textwidth]{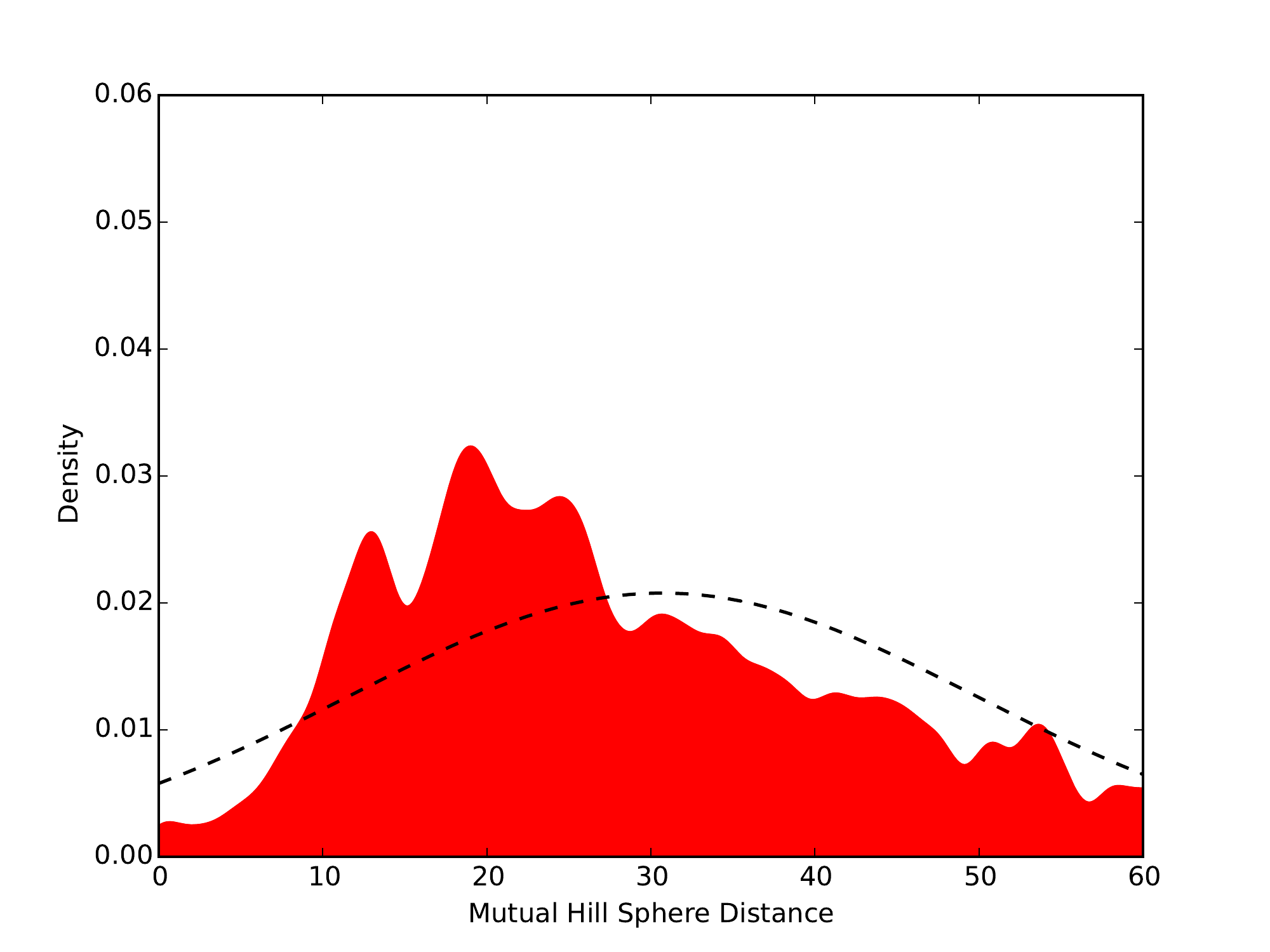}\\
\includegraphics[width=0.23\textwidth]{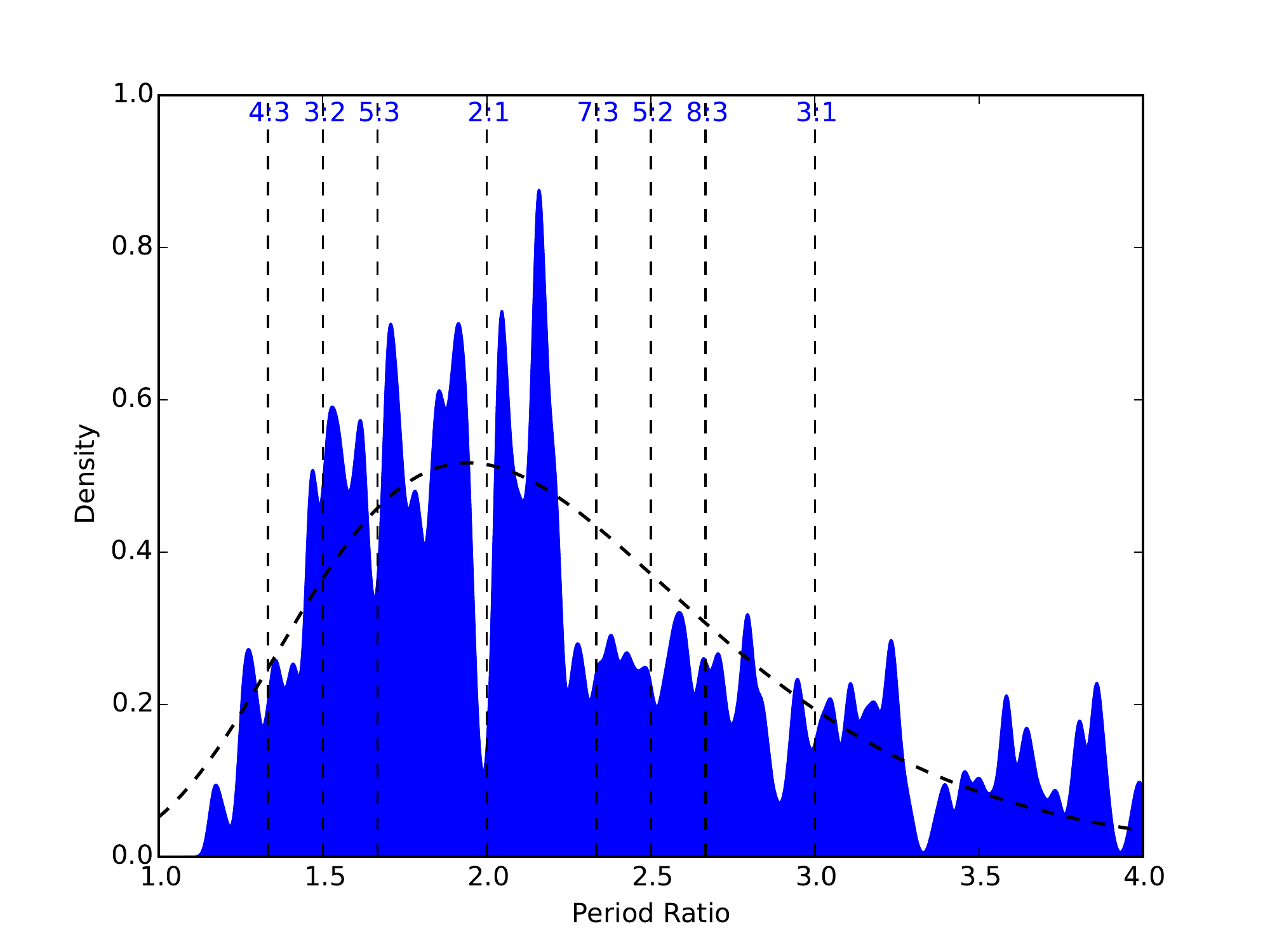} &
\includegraphics[width=0.23\textwidth]{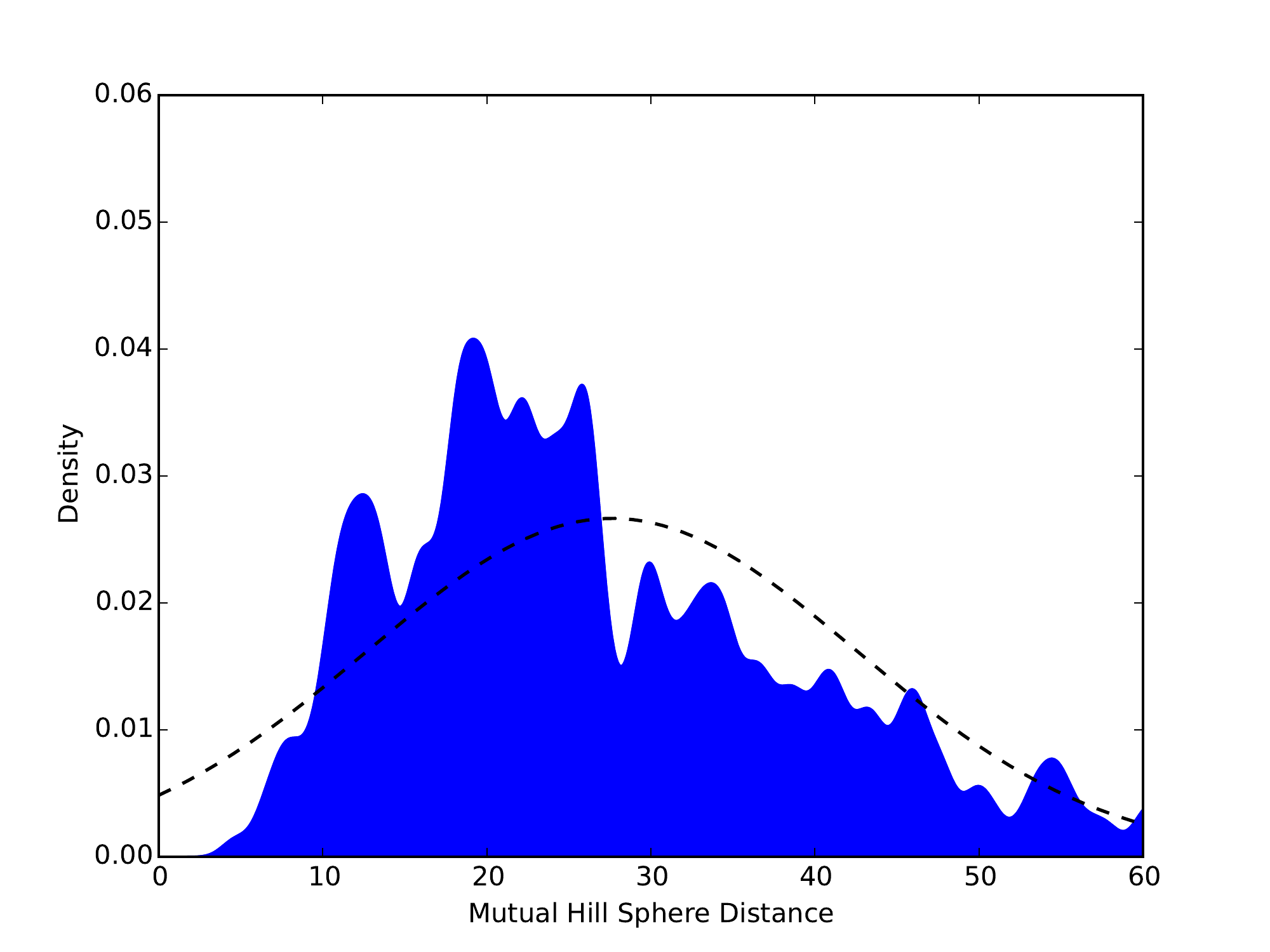}\\
\includegraphics[width=0.23\textwidth]{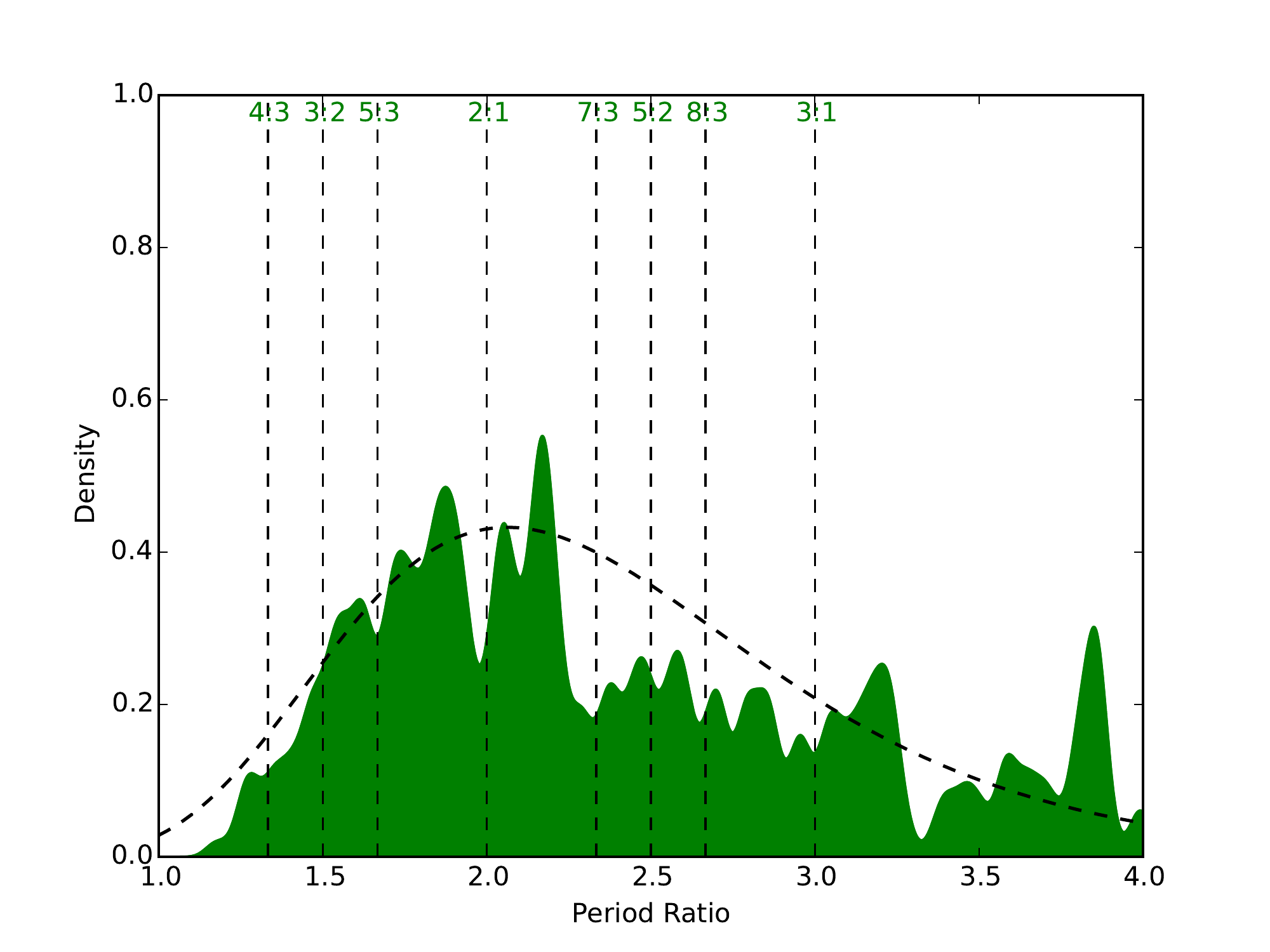} &
\includegraphics[width=0.23\textwidth]{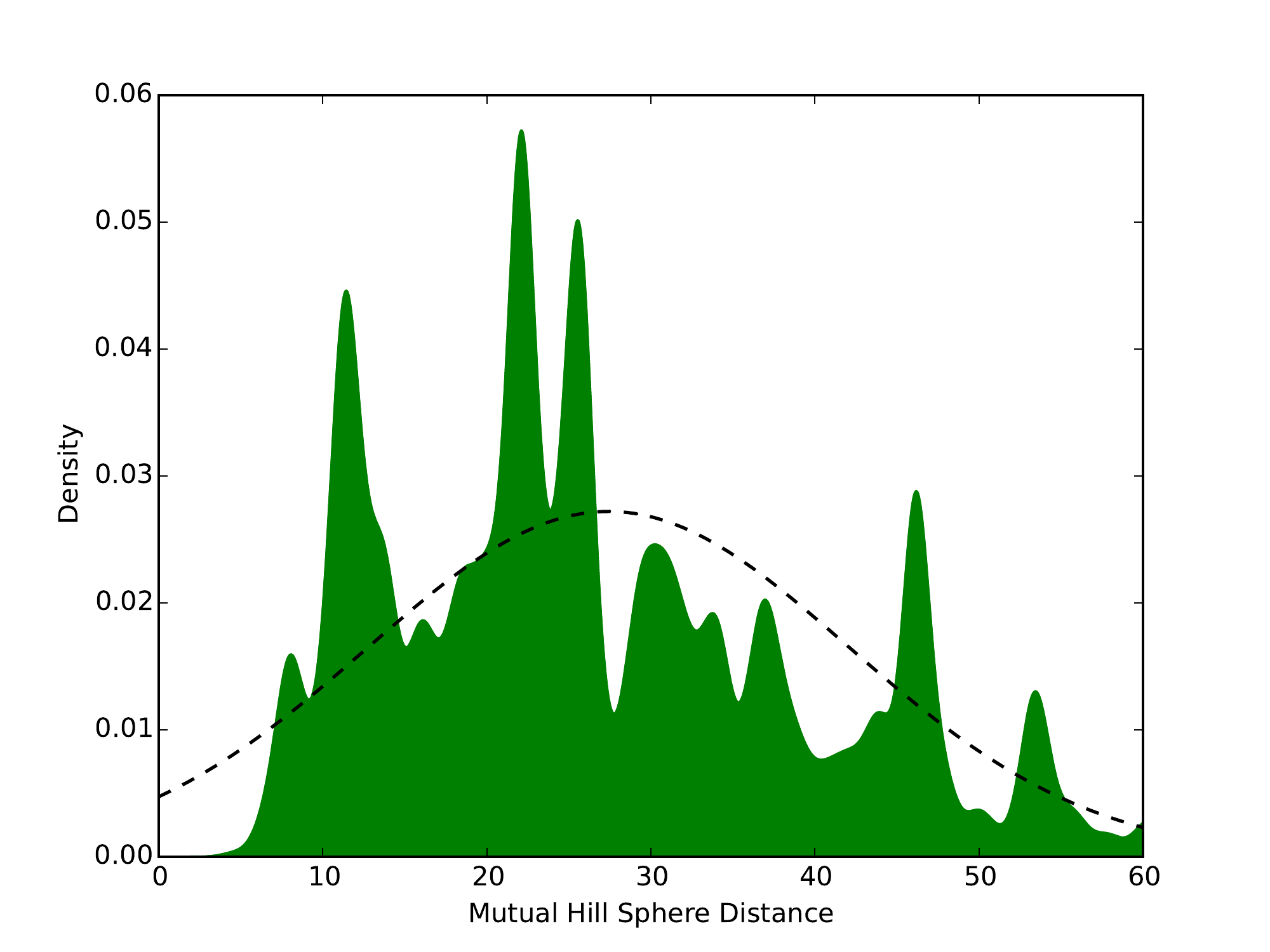}\\
\end{tabular}
\end{center}
\pagebreak
\caption{\scriptsize This series of six kernel density estimates, produced using scipy \citep{SciPy}, compares the period ratio distribution (smoothing width of $.003$) and the mutual Hill radius distribution (smoothing width of $.04$) as the \emph{Kepler} data is debiased.  The two KDEs on the upper row correspond to the raw, unweighted resonance and Hill radius distributions from the \emph{Kepler} data, removing KOI-284, KOI-351/Kepler-90, and KOI-2248. The middle row shows the unweighted resonance and Hill radius distributions after a mutual SNR cut of greater than $16$ and an impact parameter cut of less than $0.8$ is used.  The mutual SNR cut checks that the SNR cut of 16 is still maintained for each pair of planets even if the two planets' periods were swapped \citep{2013ApJ...763...41C}.  The final row shows the distribution of each statistic after each pair of planets is weighted using the inverse of the geometric probability of mutual observation, effectively removing the geometric bias.  This probability is computed using CORBITS.  The debiased resonance and Hill radius distributions were fit using the Anderson-Darling statistic to log-normal and Gaussian distributions, respectively.  The identical fit distributions are overlaid on the other histograms to demonstrate the change in the shift in distribution after the data is debiased geometrically.  Notice that the debiasing shifts the mode from the 3:2 resonance to the 2:1 resonance.}
\label{fig:per-mhs-hist}
\end{figure}

This process and the final debiased period distribution are shown in Figure \ref{fig:per-mhs-hist}. The top panel shows all period ratios, the middle panel shows period ratios after the observational cut, and the bottom panel shows the final distribution including the CORBITS geometric correction. We have confirmed that different SNR cuts and inclination distributions give very similar results. With each debiasing step, the mode of the distribution shifts slightly to longer period ratios, since short-period planets with small period ratios are the most likely to be detected. The excesses near resonances can also change somewhat in character. Also visible are large spikes due to single particularly improbable systems which should not be over-interpreted. 

We use the minimization of an Anderson-Darling statistic to identify a smooth functional fit to the cumulative distribution function with a period-ratio maximum of 4. A reasonable match is given by a log-normal distribution with $\mu = 0.82$ (${\cal P} = 2.06$) and $\sigma = 0.31$, which is shown as a dashed line in the Figure. This distribution was computed with a script written in R \citep{R}, using the libraries `ADGofTest' and `psych.' Relative to this smooth functional form are deviations due to resonances and other spikes; see \citet{2015MNRAS.448.1956S} for additional discussion on these features. 

\subsection{Mutual Hill Sphere Distribution} \label{subsec:appl-mhs}

Another metric for defining the distance between two planets is the mutual Hill sphere (MHS) distance. This is one dynamical measure of closeness or packedness \citep[e.g.,][]{1993Icar..106..247G,2011ApJS..197....8L}. Calculation of the Mutual Hill Sphere distance requires an estimate of the planetary mass; we follow \citep{2011ApJS..197....8L} and use $M=R^{2.06}$ in Earth units. Other general mass-radius relations have been proposed, but our results depend weakly on the relation used. 

With this mass estimate, we can calculate the Mutual Hill Sphere distance between neighboring pairs of planet candidates in our sample: 
$$\Delta = 2\frac{a_2 - a_1}{a_2+a_1}\left(\frac{m_1+m_2}{3M_{*}}\right)^{-1/3}$$ 
As before, we show the observed distribution, the distribution with SNR and impact parameter cuts, and the final distribution debiased using weights based on the inverse of the multi-transiting probability. Because we are considering only consecutive planets, there are 936 total pairs, of which 381 remain after the cuts. Once again, the distribution shifts slightly towards wider separations as the biases are removed. Note that this distribution draws only from neighboring pairs, so some of the larger MHS values are probably explained by a missing (non-transiting) planet in between the two observed planets. 

As before, we use the Anderson-Darling test to find a smooth functional form that approximates the fully-debiased distribution. Although the fit in this case is worse than in the period-ratio distribution, using a script written in R \citep{R}, we find a Gaussian distribution with a mean of 27.4 and a width of 14.7 to be the ``best fit.'' A few other distributions were considered, but were even less satisfactory matches.

Like the period-ratio distribution, the MHS distribution is a diagnostic of planet formation. For example, \citet{2013ApJ...775...53H} use a planet formation model of \emph{in situ} formation (migration, then assembly) to explain the detailed properties of \emph{Kepler} systems, including the MHS distribution.

CORBITS could be used to extend these analyses to multi-planet correlations, e.g., the relationship between period ratios and MHS values for the two pairs of adjacent planets in three planet subsystems. We leave this interesting extension of the above debiasing process for future work. 

\subsection{Kepler-11 and Kepler-90} \label{subsec:appl-kep11}

To demonstrate accuracy, CORBITS was also used to calculate the conditional probabilities of the existence of additional planets in the Kepler-11 system, as demonstrated in Figure 4 of \citet{2011ApJS..197....8L}. A partial reproduction of this graph is available as an example on the CORBITS GitHub repository.

A similar system is Kepler-90 (KOI-351, KIC 11442793) with 7 planets \citep{2014AJ....148...28S,2014ApJ...781...18C}. These 7 planets are not evenly spaced and, as mentioned above, produce outliers in the period ratio digram. 

\begin{figure}
\begin{center}
\includegraphics[width = 3in]{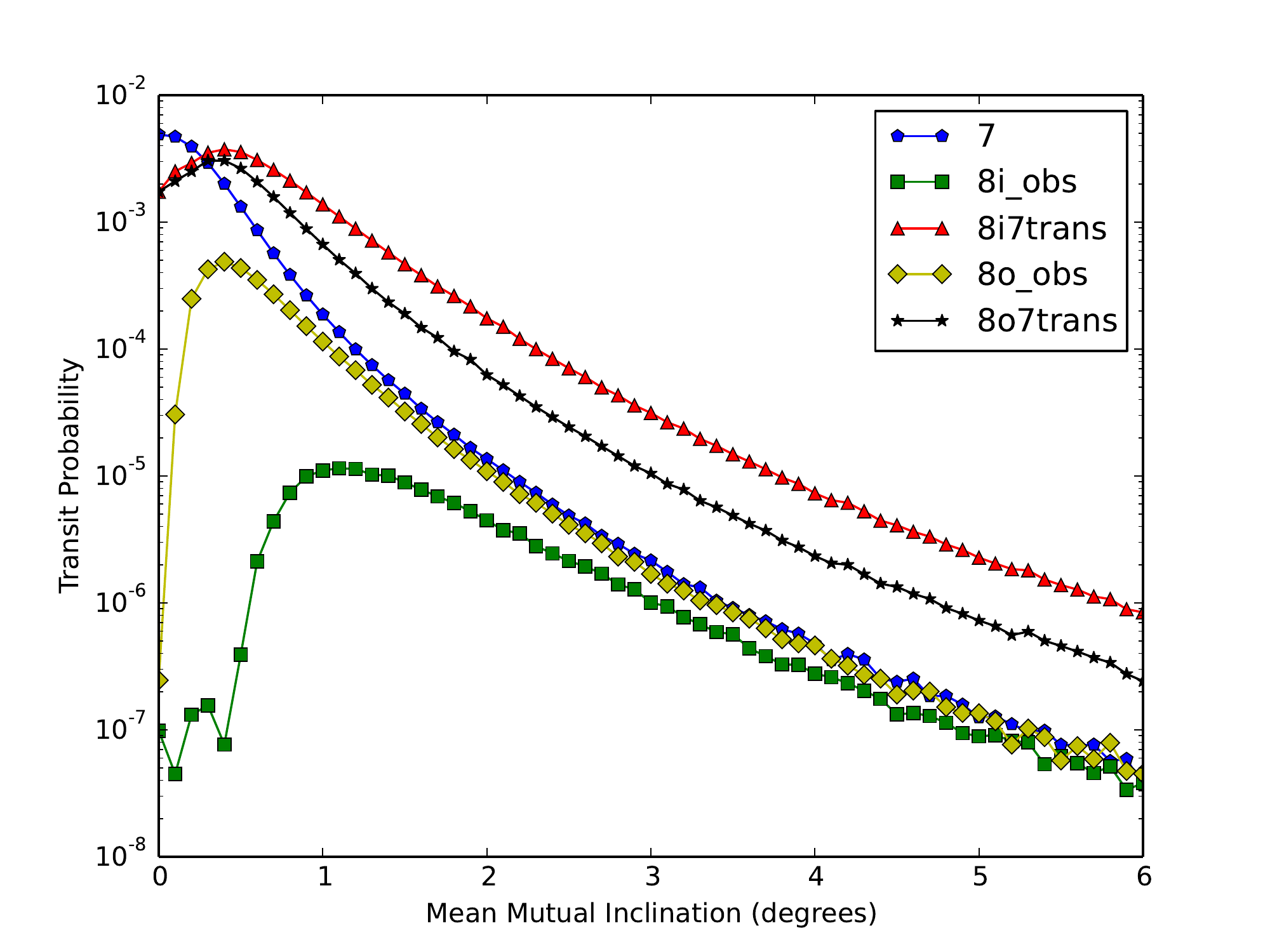}
\end{center}
\caption{Kepler-90 Multi-Transiting Probabilities as a function of Mean Mutual Inclination. Following the investigation shown in Figure 4 of \citet{2011ApJS..197....8L}, we use a combination of Monte Carlo population synthesis and CORBITS to determine the multi-transiting probability assuming different true system architectures. The legend indicates which curves are associated with which probabilities; see also the associated discussion in the text.}
\label{fig:kepler-90-probs}
\end{figure}

We begin by estimating the possible inclination distributions that could produce the observed system. Starting with the assumption that all the planet's inclinations can be drawn from a single Rayleigh distribution parametrized by a mean mutual inclination, we draw $10^5$ random systems with the observed parameters, circular orbits, and nodes drawn from a random uniform distribution. Note that the mean mutual inclination is different from the ``width'' parameter commonly used for a Rayleigh distribution by $\sqrt{\pi/2} \approx 1.25$. Each of these is studied with CORBITS to determine various probabilities as shown in Figure \ref{fig:kepler-90-probs}. 

Several different cases are considered. First, the standard case that the true Kepler-90 system has the observed 7 planets shows that the most probable case is that all 7 planets are coplanar, unsurprisingly. If the mutual inclination is relatively low, then the probability of detecting all 7 planets is  $\sim$0.001, indicating that, as \emph{Kepler} saw $\sim$1 Kepler-90 in $\sim$100,000 stars, the intrinsic frequency of such rich systems could be as high as $\sim$0.1-1\%. 

The periods of the observed planets are 7.0, 8.7, 59, 92, 125, 211, and 332 days. As with the Kepler-11 system, it is interesting to investigate whether there may be missing intermediate planets, particularly in the large gaps between the 8.7 and 59 day planets and the 125 and 211 day planets. Following the above methodology, now with 8 planets, we investigate the probabilities as a function of mean mutual inclination in the case of adding a hypothetical eighth planet with a period of 23 or 162 days. These periods are chosen as geometric means of the neighboring planets. 

If there is a 23 day planet and it is required to not transit, it must have a relatively high inclination compared to the other planets, which is rare when all 8 inclinations are drawn from the same Rayleigh distribution. If the 7 observed planets are required to transit (``8i\_obs'' in Figure \ref{fig:kepler-90-probs}) then the highest probability is $\sim$10$^{-5}$, which would suggest that nearly all \emph{Kepler} stars have systems like Kepler-90 (since $\sim$10$^{-5}$ of \emph{Kepler} targets gave KOI-351), but this seems highly unlikely. This is suggestive of the idea that there is not a missing planet (large enough to be detected in transit) between the 8.7 and 59 day planets. If, instead, we ask whether this 8 planet system would have any 7 planets transiting (not necessarily the observed 7), then the probabilities are much more reasonable (``8i7trans''), though this usually results in missing one of the two outer planets. 

If we add a 162 day planet and require that the known 7 planets are the only ones to transit (``8o\_obs''), then this limits the plausible range of mean mutual inclinations to $\sim$0.2-2 degrees. If any 7 of these planets must transit (``8o7trans''), the probabilities are similar to the nominal case. Therefore, the multi-transiting geometry is consistent with a putative non-transiting planet between the 125 and 211 day periods. 

Even when adding extra non-transiting planets, the typical mean inclination between these planets must be limited to within $\sim$3$^{\circ}$ in order for \emph{Kepler} to be lucky enough to observe 7 planets around Kepler-90. This conclusion, however, is subject to the assumption of Rayleigh inclinations and uniform nodes and other caveats discussed in $\S$\ref{subsec:MTS-caveats} and $\S$\ref{subsec:appl-sssecular}.

\begin{figure}
\begin{center}
\includegraphics[width = 3in]{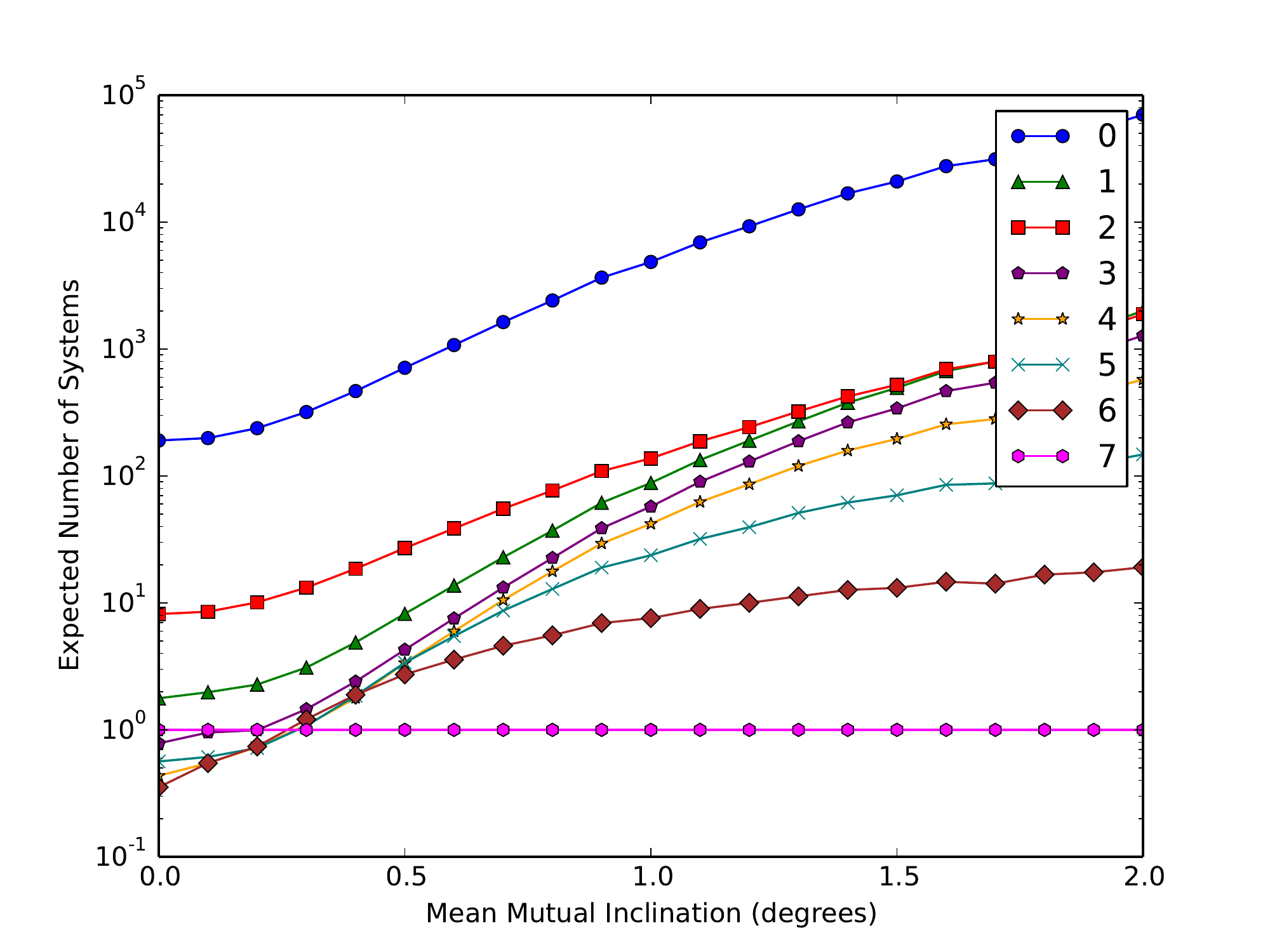}
\end{center}
\caption{Expected number of systems with $m$ transiting planets if Kepler-90 were the prototypical system architecture. Taking the Kepler-90 system and assigning inclinations from a Rayleigh distribution with a variety of mean mutual inclinations, CORBITS is used to calculate the apparent multiplicity distribution that would result. These are normalized so that the number of systems with seven transiting planet is set equal to 1. Comparison to the observed multiplicity distribution from \emph{Kepler} gives a very poor fit. For example, the number of doubly-transiting systems would either outnumber or be comparable to the number of singly-transiting systems, which is well outside the observations. This implies that the Kepler-90 system architecture is not prototypical of the STIP population. Figures 4, 5, and 6 were produced with matplotlib \citep{Hunter:2007}.}
\label{fig:kepler-90-mults}
\end{figure}

Could Kepler-90 be the ``prototype'' STIP that represents the underlying architecture of all other \emph{Kepler} STIPs? Figure \ref{fig:kepler-90-mults} shows the multiplicity distribution that would result from a population of seven-planet Kepler-90 systems as a function of mean mutual inclination. Like the previous chart, these data lines are computed using a Monte Carlo simulation aided extensively by CORBITS. For each data point, $10^4$ exoplanet systems are drawn using the same method as in the first figure. For each hypothetical system, CORBITS computes the probabilities that an observer would see 0 to 7 planets transiting. In the chart, these probabilities are normalized to show the expected number of systems with $m$-transiting planets under the assumption that exactly one 7-planet system is observed. When comparing the values generated with the observed multiplicities from \emph{Kepler}, we infer that no case is a good match. Most striking is the conclusion that a population of Kepler-90s would produce more double-transiting systems than single-transiting systems for mutual inclinations lower than 2 degrees! Similarly, there are $\sim$20 5-planet systems from \emph{Kepler}, which would imply a typical mean mutual inclination of $\sim$0.9$^{\circ}$, but this match to 5-planet systems would imply 100 2-planet systems, which is a severe underestimate. We conclude that the Kepler-90 system architecture is unlikely to merely be a fortuitously-aligned example of the remaining STIP population. 

\subsection{Known Kepler Multi-Transiting Systems} 
\label{subsec:appl-KOI}

Using CORBITS, we computed the transit probabilities of all the KOIs with at least three candidate or confirmed transiting planets and report the results in Table \ref{tbl:koi} for a variety of inclination distributions. Each trial was generated by drawing the mutual inclinations of the planets from a Rayleigh distribution with a mean mutual inclination of 0, 1, 2, or 10 degrees and the longitudes of the ascending nodes from a uniform distribution. Note that this combination of inclinations and nodes does not necessarily produce a population where the mutual inclinations are exactly drawn from a Rayleigh distribution (see RH10). 

\begin{deluxetable*}{lllllll}
\tablecaption{KOI transit probabilities}
\tablehead{\colhead{KOI} & \colhead{N} & \colhead{Planetary Periods} & \colhead{i = $0^{\circ}$} & \colhead{i = $1^{\circ}$} & \colhead{i = $2^{\circ}$} & \colhead{i = $10^{\circ}$}}
\startdata
KOI-41 & 3 & 6.89, 12.82, 35.33 & 0.016 & 0.016 & 0.014 & 0.002\\
KOI-70 & 5 & 3.70, 6.10, 10.85, 19.58, 77.61 & 0.024 & 0.021 & 0.015 & 0.000\\
KOI-82 & 5 & 5.29, 7.07, 10.31, 16.15, 27.45 & 0.042 & 0.038 & 0.029 & 0.001\\
KOI-85 & 3 & 2.15, 5.86, 8.13 & 0.041 & 0.041 & 0.041 & 0.011\\
KOI-94 & 4 & 3.74, 10.42, 22.34, 54.32 & 0.022 & 0.021 & 0.016 & 0.001\\
KOI-111 & 3 & 11.43, 23.67, 51.76 & 0.014 & 0.014 & 0.010 & 0.001\\
KOI-116 & 4 & 6.16, 13.57, 23.98, 43.84 & 0.022 & 0.022 & 0.017 & 0.001\\
KOI-117 & 4 & 3.18, 4.90, 7.96, 14.75 & 0.044 & 0.044 & 0.042 & 0.006\\
KOI-137 & 3 & 3.50, 7.64, 14.86 & 0.031 & 0.031 & 0.031 & 0.007\\
KOI-148 & 3 & 4.78, 9.67, 42.90 & 0.015 & 0.015 & 0.014 & 0.002\\ 

\enddata
\tablecomments{Table \ref{tbl:koi} is published in its entirety in the electronic edition of the Astrophysical Journal. A portion is shown here for guidance regarding its form and content. This table lists for each KOI with at least three planets, the probability of observing all the planets of each KOI transit as a function of the mean mutual inclination.  For each KOI, the probabilities were found using 1000 Monte Carlo trials.  Each trial was generated by drawing the mutual inclinations of the planets from a Rayleigh distribution with the specified mean mutual inclination and the longitudes of the ascending nodes from a uniform distribution.  The CORIBTS probability of all the planets transiting was averaged for all the trials.\label{tbl:koi}}
\end{deluxetable*}

The sharp decrease in probability from inclinations of 1$^{\circ}$ to 10$^{\circ}$ is yet another indicator that the typical mutual inclination of KOIs cannot be large. Using these simulations, we can also identify when KOIs transition between the three different regimes or Cases of multi-transiting geometry described in $\S$\ref{subsec:MTS-2}. (Recall that in Case 1, the transit region for the outer planet was completely contained within the inner planet's transit region and detecting the outer planet in transit guarantees that the inner planet's orbit will also be aligned. In Case 3, the two planets can only be seen as doubly-transiting when the observer is along the line of nodes and the multi-transiting probability is much reduced. Case 2 is an intermediate case ($\S$\ref{subsec:MTS-2}).) 

We find that the mean transition points for the known KOIs occur at mean mutual inclinations of 2.5 degrees for the transition between Case 1 and 2 and 7.0 degrees for the transition between Case 2 and 3. Assuming a Rayleigh distribution with a single width of 1.5$^{\circ}$, this suggests that Cases 1, 2, and 3 occur with frequencies of approximately $55.3\%$, $44.5\%$, and $0.2\%$, respectively. However, the system-to-system variation is very large and strongly dependent on the periods of the planets. As Case 2 is very common, there is a significant probability that we would only see one or the other planet transit, which re-emphasizes the inaccuracy of estimating the probability of the observing the entire system with the probability of observing the innermost (i.e., lowest probability) planet transit ($\S$\ref{sec:motivation}). 
    


\subsection{Exoplanet Mutual Events and Circumbinary Planets} \label{subsec:events}
A related geometric probability question is the probability that two planets will transit/occult one another as they orbit. RH10 called these ``exoplanet mutual events'' and subdivided them into two specific types: ``overlapping double transits'' or ``exosyzygies'' are mutual events that occur while both planets are transiting their parent star and ``planet-planet occultations'' are mutual events outside of transit. These mutual events are not just fun photometric oddities; they provide significant geometric information. For example, \citet{2012ApJ...759L..36H} detected an overlapping double transit in the KOI-94 system and used it to measure the otherwise difficult-to-determine true mutual inclination between the two planets in the system to within 0.55 degrees. 

In the approximation that all bodies are orbiting the system barycenter, the geometric probability governing transiting circumbinary planets (TCBPs) is analogous to a ``planet-planet occultation.'' \emph{Kepler} has discovered several TCBPs \citep[e.g.,][]{2011Sci...333.1602D}, with initial geometric interpretation discussed in \citet{2012Natur.481..475W}. The full geometric understanding of their transit probability is beyond the scope of this work, but the ``planet-planet occultation'' approximation is a possible first step. Other authors have developed useful algorithms for calculating TCBP probabilities as well \citep{2014MNRAS.437.3727K,2015MNRAS.449..781M}.  

We have used our experience with CORBITS to develop some codes investigating the geometric probabilities of exoplanet mutual events of different types. As they are tangentially related to the main CORBITS code, we discuss them in Appendix A.


\section{Conclusion} \label{sec:concl}

Even in systems with multiple known transiting planets, it is very likely that planets are missing, not only at longer periods, but an intermediate and sometimes even the shortest periods. By better understanding the biases of multi-transiting probabilities, better models can be made of the true underlying architectures of exoplanet populations.  CORBITS allows for the efficient correction of geometric bias and is essential for determining the distribution of planetary systems necessary for estimating the Fraction of Stars With Planets (FSWP), the coupled multiplicity and inclination distributions, and many other important questions. 

In Section \ref{sec:MTS}, we describe the development of CORBITS. Using the Gauss-Bonnet Theorem to find the areas of transit regions of the celestial sphere, a fast semi-analytic method is developed to determine multi-transiting probabilities. $\S$\ref{subsec:MTS-MC} establishes the accuracy and speed of CORBITS through the comparison to Monte Carlo algorithms.

As an illustration of its value and capability, we present some scientific results made possible through CORBITS in Section \ref{sec:appl}. Application to the time-evolution of the solar system is used to appreciate the complexities of multi-transiting probabilities. Turning to the \emph{Kepler} systems, CORBITS is used to geometrically debias the period ratio and mutual Hill sphere distributions, which results in shifting these towards larger values than seen in the observed distribution. Investigation of the Kepler-90 system shows it is not a typical member of the STIP population. The Appendix presents ancillary algorithms useful for understanding the probabilities of exoplanet mutual events and transiting circumbinary planets. 

Most of the results in this paper can be reproduced using the CORBITS package on GitHub at \texttt{https://github.com/jbrakensiek/CORBITS}. We hope that this resource will be helpful as the community continues to mine the information-rich vein of \emph{Kepler}'s multi-transiting systems. 

\acknowledgments
JB and DR thank Richard Rusczyk and the Art of Problem Solving for initiating their collaboration. We appreciate valuable discussions and suggestions from Eric Ford, Jared Coughlin, Matt Holman, Dan Fabrycky, Gwenael Boue, the Transit Timing Variations and Multis Working Group of the Kepler Science Team, Andrew Youdin, Jason Steffen, James Conaway, Sakhee Bhure, and others. We appreciate comments from the anonymous referee. JB thanks his parents Warren and Kathleen Brakensiek for their support. DR thanks the Institute for Theory and Computation. We thank the Division of Planetary Sciences for a Hartmann Travel Grant. Finally, we thank NASA's Origins of Solar Systems Program for supporting the conclusion of this work under grant NNX14A176G. 


\bibliographystyle{apj}
\bibliography{all}


\appendix \label{sec:appendix}

\section{Geometry of Exoplanet Mutual Events and Circumbinary Planets}

As discussed in $\S$\ref{subsec:events}, understanding the probability of exoplanet mutual events can also benefit from  geometric calculations similar to those used in the main CORBITS code. These are not included in the main GitHub distribution of CORBITS, but are available upon request from the authors. 

For the purposes of simulation, two particular phenomena are of interest (RH10).  The first, planet-planet occultations (PPOs) are the overlapping of two exoplanets from the perspective of the observer. Second, an overlapping-double transit (ODT) or exosyzygy is equivalent to a PPO when both planets are also transiting their star from the observer’s perspective.  
Our interest is to compute the probability, from the perspective of a uniformly random observer on the celestial sphere, that a PPO or an ODT occurs when the system is observed at a single uniformly random time.  We assume that the the orbital phases of the planets are independent of each other (which may be violated in resonance) so that the two planets are at independent positions in their Keplerian orbits. For this reason, we can also ignore light-travel time corrections. We can convert a randomly-chosen time to a randomly-chosen angle, using the standard method of weighting by the square of the distances of the planets from their star based on Kepler's Second Law. 

\subsection{Efficient approximation algorithm} \label{subsec:appendix-algo}

\subsubsection{Terminology}

We can approximate each of these probabilities efficiently using numerical integration.  We will take each potential position of the planets in their orbits and in each case compute what fraction of the celestial sphere can observe a PPO and an ODT, respectively.  We will let $r_1$ and $r_2$ be the distances of the two planets from the star.  Let $P_1$ and $P_2$ be the periods of the two planets.  Additionally let $R_1$ and $R_2$ be the radii of the two planets, and $R_*$ the radius of the star.  We will also let $\phi$ be the mutual inclination between the two orbits.

\subsubsection{Planet-Planet Occultation} \label{subsec:PPO}
 
Let $d(t_1,t_2)$ be the sky-projected distance between the centers of the two planets.  If we let $1$ be the radius of the celestial sphere, then the region on the celestial sphere which can observe a PPO of these planets is two diametrically opposite spherical caps of radius $$\frac{R_1+R_2}{d(t_1,t_2)}.$$  This implies that the probability of observation at that particular time is $$1-\sqrt{1-\left(\frac{R_1+R_2}{d(t_1,t_2)}\right)^2}.$$
We can integrate this probability over all positions to get the time-weighted average:
$$\frac{1}{P_1P_2}\int_{0}^{P_1}\int_{0}^{P_2}1-\sqrt{1-\left(\frac{R_1+R_2}{d(t_1,t_2)}\right)^2}\, dt_1\, dt_2.$$

This integral has been computed using numerical integration (that includes the weighting that accounts for Kepler's Second Law) via the trapezoid method with a small stepsize. As exosyzygies are generally a small fraction of parameter space, these PPO probabilities do not distinguish whether the planets are also transiting or being eclipsed. 

\subsubsection{Overlapping Double Transits} \label{subsec:ODT}

\begin{figure}
\begin{center}
\includegraphics[height = 2in]{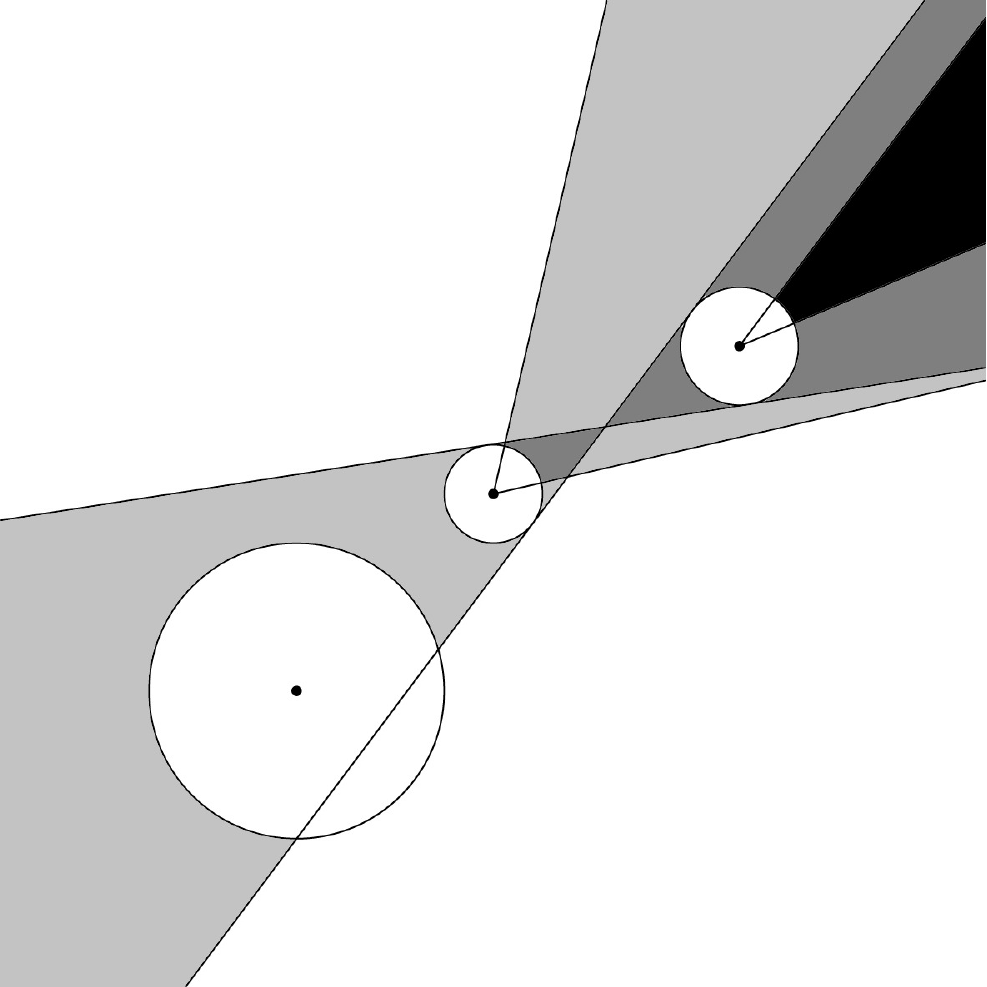}
\hfill
\includegraphics[height = 1.5in]{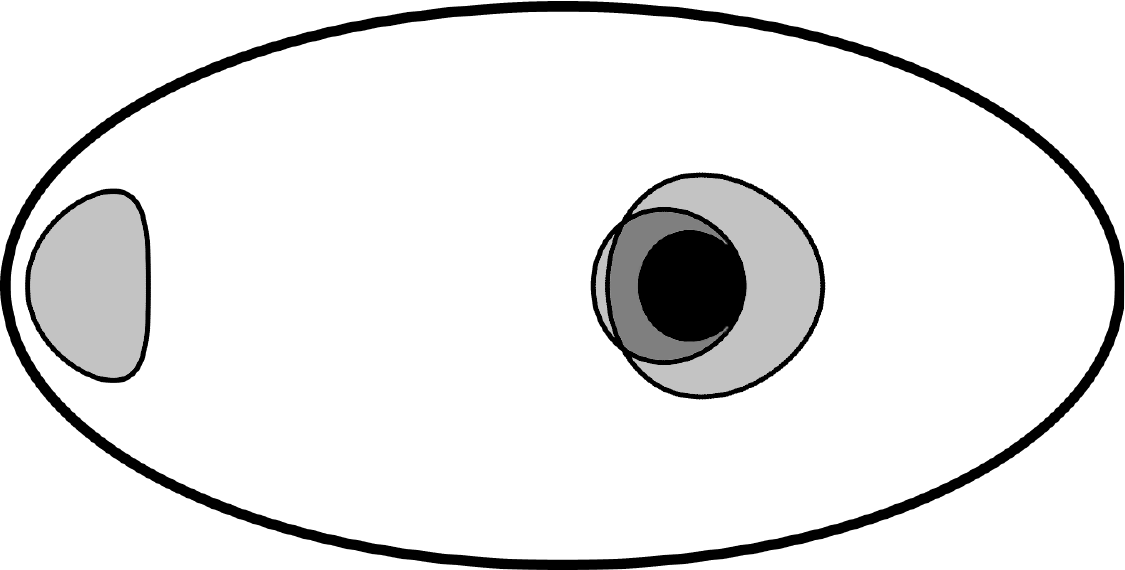}
\end{center}
\caption{On the left is a two-dimensional cross-section of two planets transiting a star.  Depending on the observer's perspective, he or she may observe (1) the PPO of the two planets, (2) the transit of the first planet, or (3) the transit of the second planet.  The region where exactly one of these events can be observed is shaded light gray, the region where exactly two of these events can be observed is shaded dark gray, and the region where all three events can be observed (an ODT) is shaded black. Note that the left light gray region corresponds to a PPO and the region of the PPO obscured by the star (double eclipse) is not indicated. On the right are the regions of the celestial sphere which can observe these events (with corresponding shading). These planets are unrealistically close to the star and each other which lead to very large regions compared to what would be seen in a typical system.}
\label{fig:ODT}
\end{figure}

In the ODT case, there are three regions of the celestial sphere to keep track of: the region where a PPO is observed and the two regions where the two planets are seen to transit at that particular moment.  In addition to the diametrically opposite spherical caps from before, there are now two single spherical caps of radii $R_*/r_1$ and $R/r_2$, respectively.  The portion of the celestial sphere which can observe an ODT is the intersection of these three regions as shown in Figure \ref{fig:ODT}.

As these regions are bounded by small circles, we can determine the area of the intersection of these regions using steps 3 to 8 of the pipeline from \S \ref{subsec:MTS-circle}.  The pipeline produces the probability for only one fixed configuration of the two planets with respect to their star.  Due to this, we numerically integrate the probabilities over $t_1$ and $t_2$.  Like for the PPO case (\S \ref{subsec:PPO}), we can approximate this integral by picking discrete values of $t_1$ and $t_2$, using the trapezoidal rule for numerical integration.

\subsection{Results} \label{subsec:appendix-results}

We performed a wide suite of Monte Carlo simulations to test the above algorithms. Our codes were compared to the Monte Carlo codes in over 6000 simulations. The deviation of the probabilities followed the normal distribution predicted by the standard error of the Monte Carlo data, indicating that our semi-analytical method is accurate. Our codes are also hundreds of times faster than the Monte Carlo equivalent.

Some qualitative relationships were found between the orbital parameters and the probabilities of PPOs and ODTs. The square of the sum of the radii of the planets was highly correlated with both the PPOs and ODTs, as would be expected. The mutual inclination and the period ratio had a complex relationship. As expected, the probabilities of ODTs and PPOs were significantly enhanced at the lowest mutual inclinations. However, the PPO probability is nearly independent of inclination, particularly when the period ratio is greater than 3.

This has implications for transiting circumbinary planets (TCBPs). While all known TCBPs have relatively low mutual inclinations between the orbit of the stellar binary and the orbit of the planet, this may be due to an observational bias. Our results here suggest that even very large mutual inclinations between the binary and planet do not diminish the probability of an occasional planet transit as was also found by \citet{2015MNRAS.449..781M} and others. Therefore, as pointed out by \citet{2014A&A...570A..91M}, this population may be revealed by the identification of single non-periodic planet transits with (spectroscopic) follow-up identifying the primary star as a non-eclipsing binary. We note that due to the large masses of the stellar components, the planetary orbit can change significantly in orientation during the course of the observations; we do not account for these important orbital changes \citep[e.g.,][]{1994P&SS...42..539S}.

Understanding the probability of finding TCBPs has the promise of identifying orbital properties of the true underlying population of circumbinary planets with implications for how these systems formed.



\end{document}